\documentclass[oneside,11pt]{article}

\usepackage[utf8]{inputenc}
\usepackage[T1]{fontenc}

\usepackage{amsthm}
\usepackage{amsmath}
\usepackage{amsfonts}
\usepackage{amssymb}

\usepackage{graphicx} 
\usepackage{float}
\usepackage{booktabs}
\usepackage{multirow}
\usepackage{rotating}
\usepackage{array}
\usepackage{xcolor}

\usepackage{nowidow}

\usepackage{breakcites}

\usepackage{enumitem}

\usepackage[top=1.5cm,bottom=2cm,right=2.5cm,left=2.5cm]{geometry}
\usepackage{titling}

\usepackage{natbib}

\usepackage{ifplatform}

\usepackage{textcomp}

\ifwindows
\fi

\ifwindows
	\usepackage{bbm}
\else
	\iflinux
		\usepackage{bbm}
	\else
		\usepackage{bbm}
	\fi
\fi

\newcommand{\lb}{\left\{}
\newcommand{\rb}{\right\}}

\newcommand{\R}{\mathbb{R}}
\newcommand{\T}{\mathbb{T}}

\newcommand{\Z}{\mathbb{Z}}

\newcommand{\rd}{\mathrm{d}}

\newcommand{\bx}{\mathbf{x}}

\newcommand{\bm}{\mathbf{m}}

\newcommand{\bX}{\mathbf{X}}

\newcommand{\bJ}{\mathbf{J}}

\newcommand{\bmu}{\boldsymbol\mu}

\newcommand{\br}{\mathbf{r}}

\newcommand{\bk}{\mathbf{k}}

\newcommand{\zero}{\mathbf{0}}

\newcommand{\bphi}{\boldsymbol\varphi}

\newcommand{\bthe}{\boldsymbol\theta}
\newcommand{\btheta}{\boldsymbol\theta}

\newcommand{\bTheta}{\boldsymbol\Theta}

\newcommand{\bSigma}{\boldsymbol\Sigma}

\newcommand{\bGamma}{\boldsymbol\Gamma}

\newcommand{\blambda}{\boldsymbol\lambda}

\newcommand{\bA}{\mathbf{A}}

\newcommand{\bI}{\mathbf{I}}
\newcommand{\bS}{\mathbf{S}}
\newcommand{\bP}{\mathbf{P}}
\newcommand{\bH}{\mathbf{H}}

\newcommand{\bM}{\mathbf{M}}
\newcommand{\bW}{\mathbf{W}}

\newcommand{\lrp}[1]{\left(#1\right)}

\newcommand{\lrb}[1]{\left\{#1\right\}}
\newcommand{\Prob}[1]{\mathbb{P}\lb #1\rb}

\newcommand{\cmod}[1]{\mathrm{cmod}\left(#1\right)}

\newcommand{\pf}[2]{\frac{\partial #1}{\partial #2}}

\newcommand{\abs}[1]{\left| #1\right|}
\newcommand{\tr}[1]{\mathrm{tr}\left[#1\right]}

\DeclareFontFamily{OT1}{pzc}{}
\DeclareFontShape{OT1}{pzc}{m}{it}{<-> s * [1.10] pzcmi7t}{}
\DeclareMathAlphabet{\mathpzc}{OT1}{pzc}{m}{it}

\usepackage[pdftex,final,bookmarksnumbered,bookmarksopen=false,breaklinks,colorlinks]{hyperref}
\hypersetup{
	final,
	bookmarksnumbered=true,
	bookmarksopen=true,
	bookmarksopenlevel=0,
	unicode=false,          
	pdftoolbar=true,        
	pdfmenubar=true,        
	pdffitwindow=false,     
	pdftitle={Toroidal diffusions and protein structure evolution},    
	pdfdisplaydoctitle=true, 
	pdftoolbar=true,
	pdfmenubar=true,
	pdflang={English},
	pdfauthor={Eduardo Garcia-Portugues, Michael Golden, Michael Sorensen, Kanti V. Mardia, Thomas Hamelryck, Jotun Hein},     
	pdfsubject={arXiv paper},   
	pdfcreator={Eduardo Garcia-Portugues},   
	pdfproducer={Eduardo Garcia-Portugues}, 
	pdfkeywords={Directional statistics}{Evolution}{Probabilistic model}{Protein structure}{Stochastic differential equation}{Wrapped normal}, 
	pdfnewwindow=true,      
	breaklinks=true,
	hidelinks,       
	linkcolor=black,          
	citecolor=black,        
	filecolor=magenta,      
	urlcolor=cyan           
}

\newtheorem{defin}{Definition}

\newtheorem{coro}{Corollary}

\newtheorem{prop}{Proposition}

\allowdisplaybreaks

\begin{document}

\title{Toroidal diffusions and protein structure evolution}
\setlength{\droptitle}{-1cm}
\predate{}%
\postdate{}%

\date{}

\author{Eduardo Garc\'ia-Portugu\'es$^{1,7}$, Michael Golden$^2$, Michael S\o{}rensen$^3$,\\ Kanti V. Mardia$^{2,4}$, Thomas Hamelryck$^{5,6}$, and Jotun Hein$^2$}

\footnotetext[1]{
	Department of Statistics, Carlos III University of Madrid (Spain).}
\footnotetext[2]{
	Department of Statistics, University of Oxford (UK).}
\footnotetext[3]{
	Department of Mathematical Sciences, University of Copenhagen (Denmark).}
\footnotetext[4]{
	Department of Mathematics, University of Leeds (UK).}
\footnotetext[5]{
	Department of Biology, University of Copenhagen (Denmark).}
\footnotetext[6]{
	Department of Computer Science, University of Copenhagen (Denmark).}
\footnotetext[7]{Corresponding author. e-mail: \href{mailto:edgarcia@est-econ.uc3m.es}{edgarcia@est-econ.uc3m.es}.}

\maketitle


\begin{abstract}
This chapter shows how toroidal diffusions are convenient methodological tools for modelling protein evolution in a probabilistic framework. The chapter addresses the construction of ergodic diffusions with stationary distributions equal to well-known directional distributions, which can be regarded as toroidal analogues of the Ornstein--Uhlenbeck process. The important challenges that arise in the estimation of the diffusion parameters require the consideration of tractable approximate likelihoods and, among the several approaches introduced, the one yielding a specific approximation to the transition density of the wrapped normal process is shown to give the best empirical performance on average. This provides the methodological building block for Evolutionary Torus Dynamic Bayesian Network (ETDBN), a hidden Markov model for protein evolution that emits a wrapped normal process and two continuous-time Markov chains per hidden state. The chapter describes the main features of ETDBN, which allows for both ``smooth'' conformational changes and ``catastrophic'' conformational jumps, and several empirical benchmarks. The insights into the relationship between sequence and structure evolution that ETDBN provides are illustrated in a case study.
\end{abstract}
\begin{flushleft}
\end{flushleft}

\section{Introduction}
\label{sec:intro}

Toroidal diffusions, this is, continuous-time Markovian processes on the torus, are useful statistical tools for modelling the evolution of a protein's backbone throughout its dihedral angles representation. This chapter reviews a class of time-reversible ergodic diffusions, which can be regarded as the toroidal analogues of the celebrated Ornstein--Uhlenbeck process, and presents their application to the construction of an evolutionary model for pairs of related proteins that aims to provide new insights into the relationship between protein sequence and structure evolution. \\

The chapter is organized as follows. The rest of this section provides a brief background on protein structure and protein evolution, while it outlines the fundamentals of ETDBN (standing for Evolutionary Torus Dynamic Bayesian Network), a probabilistic model for protein evolution. Section \ref{sec:tordiff} studies toroidal diffusions, the main methodological innovation behind ETDBN. The important challenges that arise in the estimation of the diffusion parameters require the consideration of tractable approximate likelihoods and, among the several approaches, the one yielding a specific approximation to the transition density of the wrapped normal process is shown to give the best empirical performance on average. ETDBN is described in detail in Section \ref{sec:etdbn}: its structure as a hidden Markov model featuring a wrapped normal process and two continuous-time Markov chains, its training from real data, and its empirical performance in several benchmarks. A distinctive feature of ETDBN is that it allows for both ``smooth'' and ``catastrophic'' conformational changes on the protein structure evolution by combining two evolutionary regimes within each hidden node. In addition, ETDBN provides new insights into the relationship between sequence and structure evolution through the analysis of hidden states. These two points are thoroughly illustrated in the case study given in Section \ref{sec:casestudy}.

\subsection{Protein structure}
\label{subsec:prots}

Proteins are large and complex biomolecules that are vital to all forms of life -- from virus to human \citep{dill1999polymer,dill2012protein}. Their main functions include defence against infections, catalysing chemical reactions, transfer of signals between cells, transport of other molecules such as oxygen, and providing structure and support for cells, tissues and organs. Chemically, proteins are simply linear polymers of amino acids, which for most proteins fall into $20$ different types. \\

The amino acid sequence of a protein is encoded in the DNA of its matching gene, and is easy to obtain experimentally. However, most proteins also adopt a specific three-dimensional shape, which is the result of a folding process in which the linear polymer folds into a compact shape. This process is driven by the so-called hydrophobic effect -- fatty amino acids repel water and become buried in the so-called hydrophobic core at the centre of the protein \citep{dill1999polymer}. The three-dimensional shape of a protein is often crucial for its function. Unfortunately, unlike a protein's amino acid sequence, a protein's three-dimensional structure is hard to obtain, requiring expensive and elaborate experimental techniques such as X-ray crystallography or nuclear magnetic resonance. Therefore, computational techniques to predict the three-dimensional structure of proteins are in high demand and an active area of research \citep{dill2012protein}. Currently, most structure prediction methods are essentially heuristic in character, but sophisticated probabilistic models of protein structure and sequence are increasingly being developed, applied, and accepted \citep{Boomsma2008,valentin2013,Marks:2011kx}. 

\begin{figure}[h!]
	\centering
	\includegraphics[width=0.75\textwidth, clip, trim=0cm 0.4cm 0cm 0.2cm]{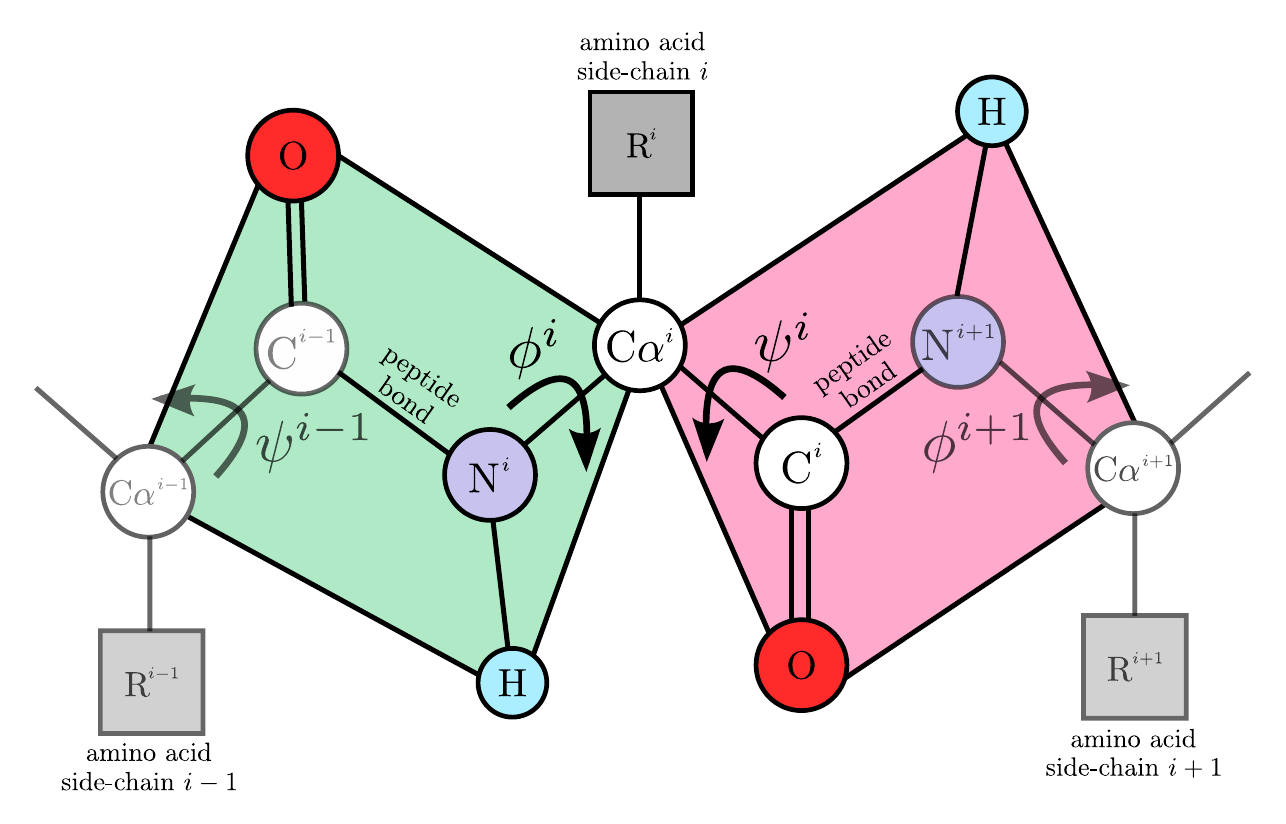}
	\caption{\small Small section of the protein backbone showing the $\phi$ and $\psi$ dihedral angles between the planar atomic configurations.}%
	\label{fig:dihedralangles}%
\end{figure}

From a geometrical point of view, a protein's shape can be fully specified by a set of bond lengths between atom pairs, bond angles between three connected atoms, and \textit{dihedral angles} between four connected atoms. The $\phi$ (specified by C$^{i-1}$-N$^{i}$-C$\alpha^{i}$-C$^{i}$ atoms) and $\psi$ (N$^{i}$-C$\alpha^{i}$-C$^{i}$-N$^{i+1}$) dihedral angles describe the part that is common to all amino acids and they form the most important degrees of freedom of the protein's structure, together with the $\omega$ dihedral angle which describes the configuration of the planar peptide bond between consecutive amino acids. The $\omega$ angle is unusual in the sense that it adopts values very close to either $0^\circ$ or $180^\circ$, rather than a continuous range of values as is the case for $\phi$ and $\psi$. As the great majority of $\omega$ angles are close to $0^\circ$, this angle is often not a crucial degree of freedom. The $(\phi,\psi)$ dihedral angles essentially describe the overall curve of the linear part of the protein polymer, namely the protein's backbone structure. In addition, each amino acid also has a variable side-chain, which can contain up to four dihedral angles. The dihedral angles of the side-chains (grey boxes in Figure \ref{fig:dihedralangles}) can be fairly well predicted given the $(\phi,\psi)$ angles, and are thus a less critical degree of freedom. \\

A benefit of the dihedral representation is that it bypasses the need for \textit{structural alignment}, which is the process of rotating and translating the three-dimensional Cartesian coordinates of the protein structures to closely align their atoms. Protein structure models on Cartesian coordinates require a structural alignment of proteins to compare them \citep{herman2014simultaneous}, a potential source of uncertainty. In addition, the dihedral representation introduces a simple distance between two conformations of the same protein: the average of the \textit{angular distances} (see \citet{Golden:evo} for further details) between pairs of dihedral angles at each amino acid. This distance is a computationally faster alternative to the Root Mean Squared Deviation (RMSD) between the structurally aligned three-dimensional atomic coordinates of two proteins. \\

The structure of a protein can be further used to label local regions of a protein by their \textit{secondary structure}. Secondary structure is a coarse-grained description of protein structure where each amino acid residue in a protein is assigned a label associated to the main structural motif which they belong to. Hence, knowledge of the secondary structure notably constrains the conformational possibilities of the dihedral angles. The two most important secondary structure classes are $\alpha$-helices and $\beta$-sheets (see the cartoon depictions of helices and flat arrows, respectively, in Figure~\ref{fig:phylogeny}). ETDBN parametrizes a protein as a discrete sequence of amino acids ($\bA$), a continuous sequence of dihedral angles ($\bX$), and a discrete sequence of secondary structure labels ($\bS$). More precisely, a protein comprised of $n$ amino acids is encoded mathematically as $\bP:=(\bA,\bX,\bS)\equiv\big(\bP^1,\ldots,\bP^n\big)$, where $\bP^i:=(A^i, \bX^i, S^i)$, $\bA:=(A^1,\ldots,A^n)$, $\bX:=(\bX^1,\ldots,\bX^n)$, $\bS:=(S^1,\ldots,S^n)$, and $\bX^i:=\langle\phi^i,\psi^i\rangle$, $i=1,\ldots,n$. This notation is used extensively in Section\nolinebreak[4] \ref{sec:etdbn}.

\subsection{Protein evolution}
\label{subsec:evolution}

\begin{figure}[!h]
	\centering
	\includegraphics[width=0.75\textwidth]{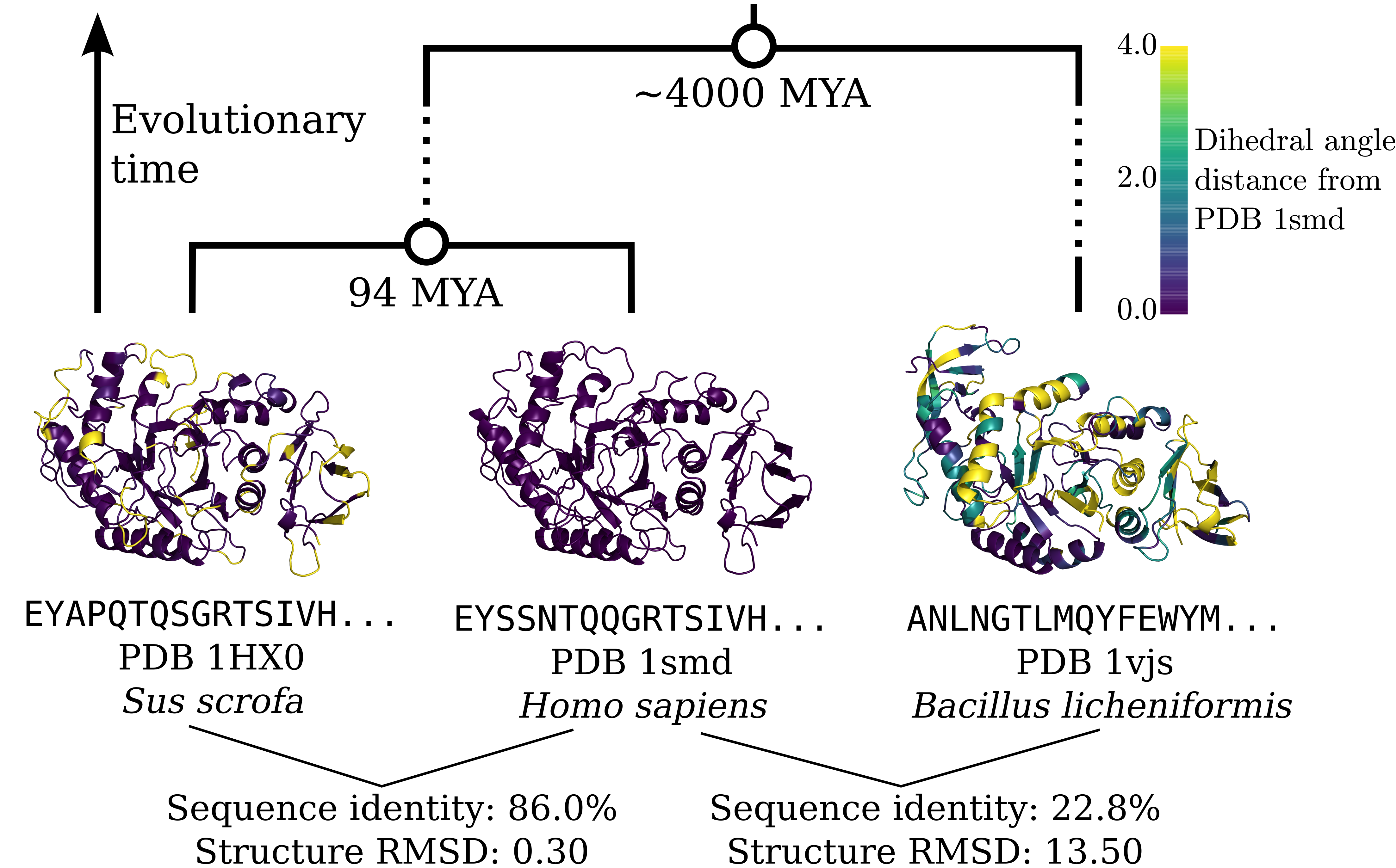}
	\caption{\small A phylogenetic tree relating three homologous amylase proteins, comparing their sequence identity and structural divergence. Overlaid on the cartoon representations of PDB 1hx0 (left) and PDB 1vjs (right) are the dihedral angle distances compared to PDB 1smd (centre) at each aligned amino acid position. The degree of structural divergence (reported as ``Structure RMSD'') is measured using the RMSD between the aligned atomic coordinates of the corresponding PDB (Protein Data Bank; \citet{joosten2011series}) files. The wild boar (\textit{Sus scrofa}) and human (\textit{Homo sapiens}) amylases share an ancestor 94 Million Years Ago (MYA). The third amalyse from \textit{Bacillus licheniformis} shares a common ancestor with the other two species approximately 4000 MYA. The common (unobserved) ancestors are represented by white nodes. Notice that the dependencies are weaker the further back in time two proteins share an ancestor.}%
	\label{fig:phylogeny}%
\end{figure}

Two or more proteins are termed \textit{homologous} if they share a common ancestor. A descendant protein is assumed to diverge from ancestral proteins via a process of mutation. Multiple homologous proteins that have diverged from a common ancestor will have dependencies (similarities) in their sequence and structure. These dependencies can be represented by a phylogenetic tree (see Figure~\ref{fig:phylogeny}). One way in which these evolutionary dependencies manifest themselves is in the degree of amino acid sequence similarity shared amongst the homologous proteins. Their strength is assumed to be a result of two major factors: the time since the common ancestor and the rate of evolution. \\

Failing to account for evolutionary dependencies can lead to misleading inferences \citep{felsenstein1985phylogenies}. For example, strong signals of structural conservation may be wrongly attributed to the selective maintenance of structural features due to their supposed functional importance. In reality, these structural similarities may simply be due to the close evolutionary relatedness of the proteins being analysed. On the other hand, accounting for evolutionary dependencies allows information from homologous proteins to be incorporated in a principled manner. This can lead to more accurate inferences, such as the prediction of a protein structure from a homologous protein sequence and structure, known as \textit{homology modelling} \citep{arnold2006swiss}. Whilst stochastic models do not yet out-perform standard homology modelling approaches in terms of predictive accuracy, they provide a statistical foundation that allows for evolutionary parameters and their associated uncertainties to be estimated in a rigorous manner.

\begin{figure}[!h]
	\centering
	\includegraphics[width=0.75\textwidth]{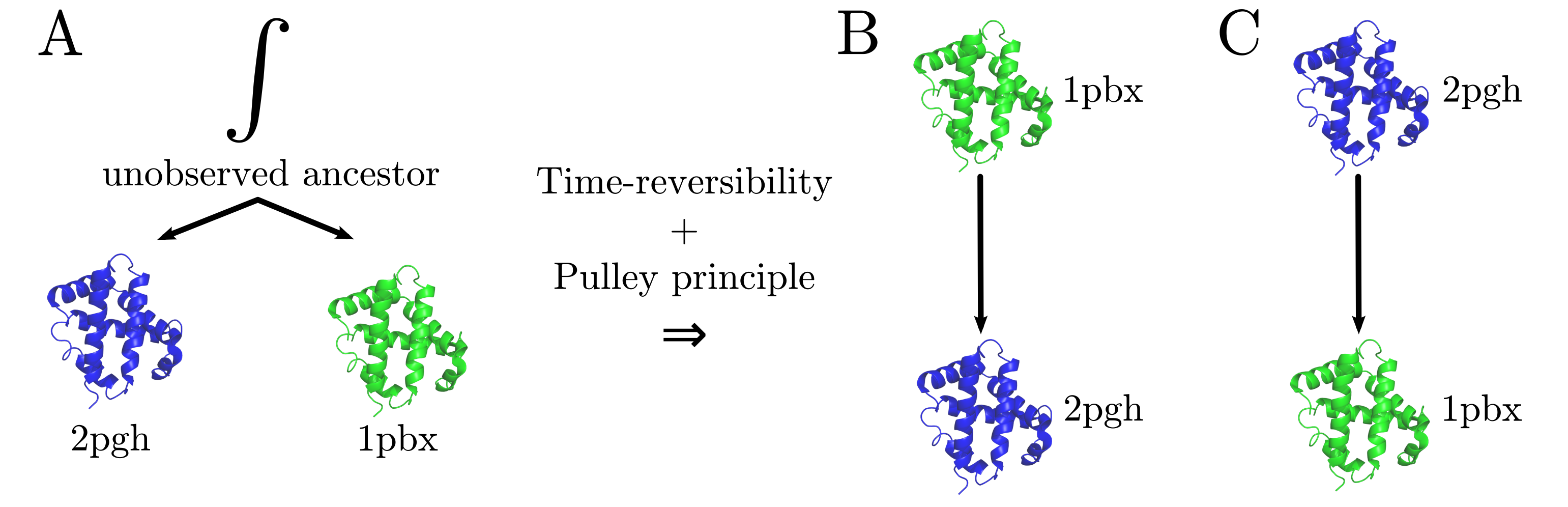}
	\caption{\small Illustration of the pulley principle in a pairwise phylogeny. To calculate the likelihood of two homologous proteins it is usually necessary to integrate out the sequence and structure of the unobserved ancestor at the root of the tree (A). However, under a time-reversible stochastic process, the pulley principle (\textit{i.e.}, assuming that the protein at the root of the tree is drawn from the stationary distribution of the process) allows any of the two observed proteins to be set as the root (B or C) without altering the likelihood. This has the advantage of avoiding a costly integration.}%
	\label{fig:reversibility}%
\end{figure}

A common property of the stochastic processes used to model molecular evolution is \textit{time-reversibility}. Whilst biological processes are not expected to obey time-reversibility, it is nevertheless considered a reasonable assumption that provides computational advantages. For example, in a pairwise evolutionary model, such as ETDBN, time-reversibility together with the \textit{pulley principle} \citep{felsenstein1981evolutionary} allows the phylogenetic tree to be re-rooted on an observed protein (where the root is assumed to be drawn from the stationary distribution of the stochastic process), avoiding a costly integration over an unobserved ancestral protein (see Figure~\ref{fig:reversibility} for an illustration). \\

Two homologous proteins can not only differ in amino acid identity due to mutations that have occurred since their common ancestor, but also by the insertion and deletion of amino acids. Insertions and deletions are collectively referred to as \textit{indels}. Accounting for this in practice involves a \textit{sequence alignment} of both proteins. A sequence alignment is constructed by attempting to identify positions that have evolved via mutation alone (termed \textit{homologous} positions) and positions that have incurred indels. Gap characters (typically denoted by a `$-$') are inserted in either sequence at indel positions such that both sequences with newly inserted gaps have the same length, thus forming a sequence alignment. Each position in this sequence alignment is referred to as an \textit{aligned site}. Section \ref{sec:etdbn} makes extensive use of this terminology.

\subsection{Towards a generative model of protein evolution}
\label{subsec:generative}

A key step in modelling protein structure evolution is selecting a structural representation and an adequate stochastic process. The first investigations of protein structure evolution represented protein structure using the three-dimensional Cartesian coordinates of protein backbone atoms and employed diffusions processes to model the relationship between structural distance and sequence similarity \citep{gutin1994evolution,grishin1997estimation}. More recent publications \citep{challis2012stochastic, herman2014simultaneous} likewise used the three-dimensional Cartesian coordinates of amino acid C$\alpha$ atoms to represent protein structure together with Ornstein--Uhlenbeck (OU) diffusions to construct Bayesian probabilistic models of protein structure evolution. These models emphasise estimation of evolutionary parameters such as the evolutionary time between species, tree topologies and sequence alignment, and attempt to fully account for sources of uncertainty. For the sake of computational tractability, the aforementioned approaches treat the Cartesian coordinates as evolving independently. From a generative perspective, the Gaussian-like and independence assumptions on the evolution of the C$\alpha$ atoms will lead to evolved proteins with C$\alpha$ atoms that are unnaturally dispersed in space. \\

Rather than using a Cartesian coordinate representation, ETDBN \citep{Golden:evo} uses a dihedral angle representation of protein structures motivated by the non-evolutionary TorusDBN model \citep{Boomsma2008, boomsma2014equilibrium}. TorusDBN represents a single protein structure as a sequence of dihedral angle pairs, which are modelled using bivariate von Mises distributions \citep{Mardia2012a}. The evolution, rather than the distribution, of dihedral angles in ETDBN is modelled using a novel diffusion process \citep{Garcia-Portugues:lang} aimed to provide a more realistic tool for capturing the evolution of the underlying protein structure manifold. This diffusive process is coupled with two continuous-time Markov chains that model the evolution of amino acids and secondary structures labels. An additional coupling is introduced such that an amino acid change can lead to a \textit{jump} in dihedral angles and a change in diffusion process, allowing the model to capture changes in amino acid that are directionally coupled with changes in dihedral angle or secondary structure. As in \citet{challis2012stochastic} and \citet{herman2014simultaneous}, the indel evolutionary process is also modelled to account for sequence alignment uncertainty by summing over all possible histories of insertions and deletions using a birth-death process as a prior \citep{thorne1992inching}. \\

For computational expediency, in the development of ETDBN it was key that the toroidal diffusion used to model the evolution of dihedral angles was time-reversible and allowed for a tractable likelihood approximation for arbitrary times. Additionally, it was desirable to efficiently sample dihedral angles under the diffusion to perform inference. The next section outlines in detail the development of a diffusion meeting the above criteria, which was guided by the goal of finding a toroidal OU-like process. The treatment in detail of ETDBN is therefore postponed until Section \ref{sec:etdbn}.

\section{Toroidal diffusions}
\label{sec:tordiff}

The three-dimensional backbone of a protein comprised by $n$ amino acids can be described as a sequence of $n-2$ pairs of dihedral angles $\{(\phi,\psi)_i\}_{i=1}^{n-2}$ (the first and last pairs are often disregarded due to missing $\phi$ and $\psi$ angles, respectively). Therefore, a statistical tool for modelling the evolution of a protein's backbone using its true degrees of freedom is a continuous-time stochastic process on the torus $\T^p=[-\pi,\pi)\times\overset{p}{\cdots}\times[-\pi,\pi)$ (with $-\pi$ and $\pi$ identified), with $p=n-2$ or $p=2$, depending on whether the backbone is modelled as a whole or piecewisely as a combination of pairs of dihedral angles, respectively. \\

One of the first continuous-time processes on the circle ($p=1$) was proposed by \citet{Kent1975} as the solution to the Stochastic Differential Equation (SDE)
\begin{align}
\rd X_t=\alpha\sin(\mu-X_t)\rd t+\sigma\rd W_t,\label{eq:vMOU}
\end{align}
where $\{W_t\}$ is a Wiener process, $\alpha>0$ is the drift strength, $\mu\in[-\pi,\pi)$ is the circular mean, and $\sigma>0$ is the diffusion coefficient. This process, termed as the von Mises process, is attracted to $\mu$, with a drift approximately linear in the neighbourhood of $\mu$. The process is ergodic and its stationary distribution (sdi) is a $\mathrm{vM}\big(\mu,
\frac{2\alpha}{\sigma^2}\big)$, the von Mises (vM) distribution with mean $\mu$ and concentration $\frac{2\alpha}{\sigma^2}$, usually regarded as a circular analogue of the Gaussian distribution. The similarities of \eqref{eq:vMOU} with the celebrated OU process
\begin{align}
\rd X_t=\alpha(\mu-X_t)\rd t+\sigma\rd W_t,\label{eq:OU}
\end{align}
whose sdi is a $\mathcal{N}\big(\mu,\frac{\sigma^2}{2\alpha}\big)$, supported \citet{Kent1975}'s claim about the vM process being ``the circular analogue of the OU process on the line''. \\

Despite the similarities of \eqref{eq:vMOU} and \eqref{eq:OU}, only the latter presents a closed-form analytical expression for its transition probability density (tpd), thus making its maximum likelihood inference fully tractable. The unavailability of the tpd is usually the case for the majority of \textit{diffusions}, the continuous-time Markovian processes solving SDEs. In a general setting, the tpd of the $p$-dimensional Euclidean diffusion
\begin{align}
\rd \bX_t=b(\bX_t)\rd t+\sigma(\bX_t)\rd\bW_t, \label{eq:sde}
\end{align}
where $b:\R^{p}\rightarrow\R^{p}$ is the drift function, $\sigma:\R^{p}\rightarrow\R^{p\times p}$ is the diffusion coefficient, and $\bW_t=(W_{t,1},\ldots,W_{t,p})'$ is a vector of $p$ independent standard Wiener processes ($'$ denotes transposition), is denoted as $p_t(\cdot\, | \,\bx_s)$. It represents the density function of the conditional distribution of $\bX_{t+s}$ given $\bX_s=\bx_s$. The tpd is only given implicitly as the solution to the Fokker--Planck equation, this is, the Partial Differential Equation (PDE)
\begin{align}
\pf{}{t} p_t(\bx\, | \,\bx_s)=&\,-\sum_{i=1}^p\frac{\partial}{\partial
	x_i} (b_i(\bx)p_t(\bx\,|\,\bx_s))\notag\\
&+
\frac{1}{2}\sum_{i,j=1}^p\frac{\partial^2}{\partial x_i\partial x_j}
(V_{ij}(\bx)p_t(\bx\, | \,\bx_s)), \label{eq:pde}
\end{align}
with $\bx,\bx_s\in\R^p$, $V(\cdot):=\sigma(\cdot)\sigma(\cdot)'$, and initial condition $p_0(\bx\, | \,\bx_s)=\delta(\bx-\bx_s)$ ($\delta(\cdot)$ represents Dirac's delta). This PDE has no explicit solution except for very few particular choices of $b$ (e.g., linear) and $V$ (e.g., constant). \\

Defining diffusive processes $\lrb{\bTheta_t}$ whose state space is $\T^p$, such as \eqref{eq:vMOU}, requires certain caution for achieving proper transitions of the process through the identified points $-\pi$ and $\pi$. A useful construction consists in regarding $\lrb{\bTheta_t}$ as a Euclidean process $\lrb{\bX_t}$ that is \textit{wrapped} into its principal angles by the \textit{wrapping operator} $\cmod{\cdot}:=((\cdot+\pi)\mod 2\pi)-\pi$. \\\nowidow[3]

\begin{defin}[Toroidal diffusion]
	\label{def:tordi}
	The stochastic process $\{\bTheta_t\}\subset\T^p$ is said to be a \emph{toroidal diffusion} if it arises as the wrapping $\bTheta_t=\cmod{\bX_t}$ of the diffusion \eqref{eq:sde}, and if $b$ and $\sigma$ are $2\pi$-periodical:
	\[
	b(\bx+2\bk\pi)=b(\bx),\quad\sigma(\bx+2\bk\pi)=\sigma(\bx),\quad\forall\bk\in\Z^p,\,\forall\bx\in\R^p.
	\]
	The toroidal diffusion coming from the wrapping of \eqref{eq:sde} is denoted as $\rd \bTheta_t=b(\bTheta_t)\rd t+\sigma(\bTheta_t)\rd \bW_t$.
\end{defin}

\begin{figure}[h!]
	\centering
	\includegraphics[width=0.75\textwidth]{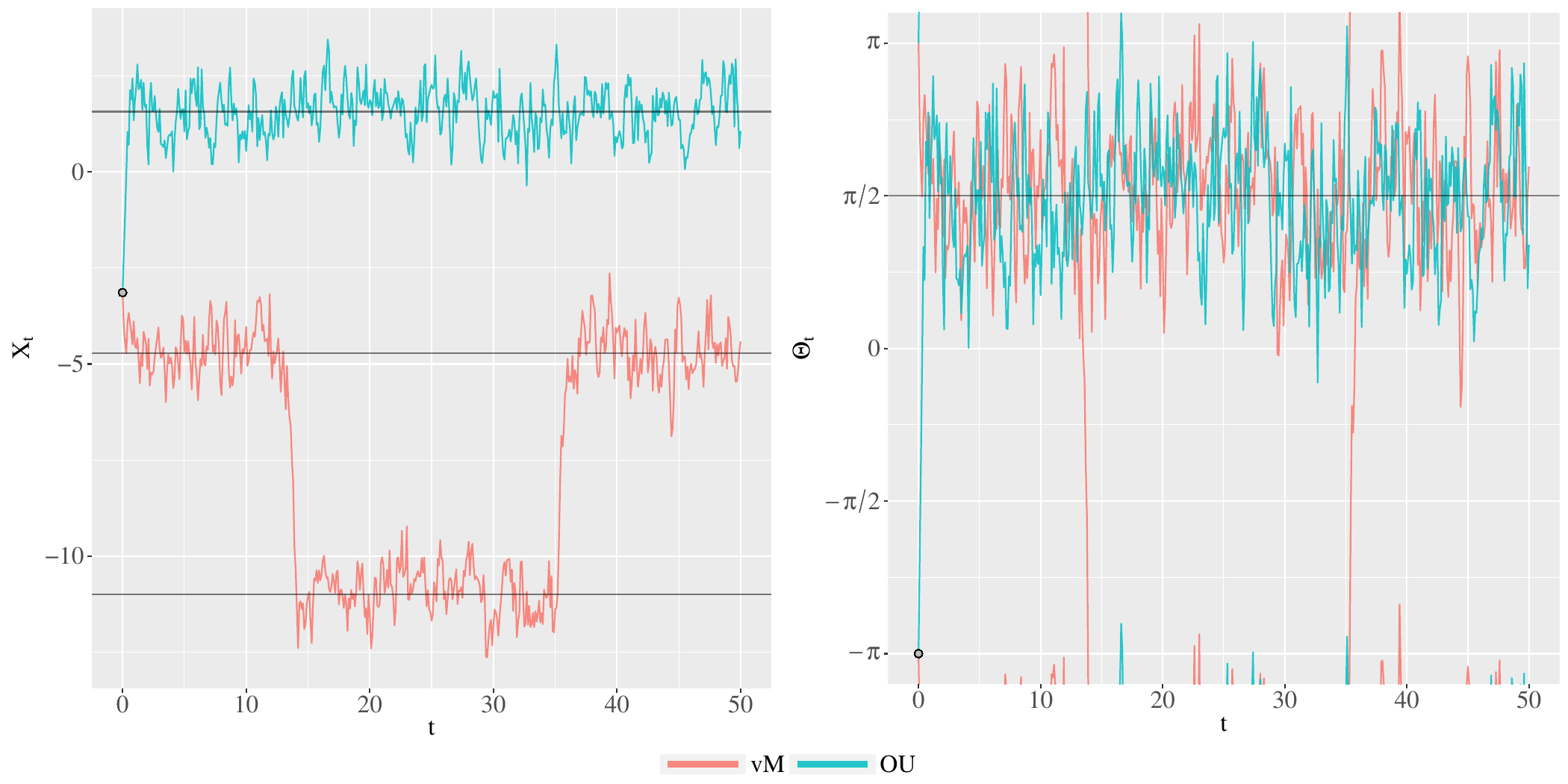}
	\caption{\small Trajectories of the vM and OU processes on the circle for $t\in[0,50]$, $\alpha=5$, $\mu=\frac{\pi}{2}$, $\sigma=2$, and starting point $x_0=-\pi$. The trajectories in the left correspond to the unwrapped processes $\{X_t\}$, whereas the ones in the right are their wrapped versions $\{\Theta_t\}$. The vM process is not ergodic in $\R$ but its wrapping is ergodic in $[-\pi,\pi)$. The OU and WOU processes are ergodic in $\R$ and $[-\pi,\pi)$, respectively. However, the WOU process is \textit{not} Markovian and is not a toroidal diffusion. Note that only the vM process is able to travel in both directions towards $\mu$.\label{fig:diff}}
\end{figure}

With the previous definition and notation, for a given $t$, $\bX_t=\bTheta_t+2\mathrm{wind}(\bX_{t})\pi$, where $\mathrm{wind}(\bX_{t}):=\lfloor\frac{\bX_{t}+\pi}{2\pi}\rfloor\in\Z^p$ is the \emph{winding number} of $\bX_t$. The fact that $b$ and $\sigma$ are required to be periodic implies that $\{\bX_t\}$ is a non-ergodic process in $\R^p$ and that $\{\bTheta_t\}$ is a Markovian process. Otherwise $\bTheta_{t_2}\,|\,(\bTheta_{t_1}, \, \bTheta_{t_0})$, with $t_2>t_1>t_0$, would not depend only on $\bTheta_{t_1}$ but also on $\mathrm{wind}(\bX_{t_1})$ in $\bX_{t_1}=\bTheta_{t_1}+2\mathrm{wind}(\bX_{t_1})\pi$. The construction of the vM diffusion and the Wrapped OU (WOU) process on the circle is illustrated in Figure \ref{fig:diff}. 

In the rest of this section we illustrate how to construct OU-like toroidal diffusions and focus on the particular bivariate diffusion we employ in ETDBN, whose properties and approximate inference are analysed.

\subsection{Toroidal Ornstein--Uhlenbeck analogues}
\label{subsec:torou}

Let $f$ be a probability density function (pdf) over $\R^p$. The so-called \textit{Langevin diffusions} are the family of diffusions \eqref{eq:sde} such that the $i=1,\ldots,p$ entries of the drift are given by
\begin{align}
b_i(\bx)=\,&\frac{1}{2}\sum_{j=1}^pV_{ij}(\bx)\pf{}{x_j}\log f(\bx)\notag\\
&+\det V(\bx)^\frac{1}{2}\sum_{j=1}^p\pf{}{x_j}\lrp{V_{ij}(\bx)\det V(\bx)^{-\frac{1}{2}}}.\label{eq:lange}
\end{align}
If $V(\bx)=\sigma(\bx)\sigma(\bx)'=\bSigma$ with $\bSigma$ a covariance matrix, \textit{i.e.} if the diffusion coefficient is constant, then the Langevin diffusions are of the form
\begin{align}
\rd \bX_t=\frac{1}{2}\bSigma\nabla\log f(\bX_t)\rd t+\bSigma^\frac{1}{2}\rd\bW_t,\label{eq:langcte}
\end{align}
where $\nabla$ denotes the gradient operator. Under mild regularity conditions on $f$ and $\sigma$, these diffusions are ergodic with stationary density $f$. The construction of Langevin toroidal diffusions is achieved by wrappings of Langevin diffusions, imposing now that $f$ is a toroidal density: $\int_{\T^p}f(\btheta)\rd\btheta=1$ and $f(\btheta+2\bk\pi)=f(\btheta)$ $\forall\btheta\in\T^p,\,\bk\in\Z^p$. The following result guarantees that the stationary density of such toroidal diffusion is indeed $f$, with the addition of a highly convenient characterization.

\begin{prop}[\citet{Garcia-Portugues:lang}]
	\label{prop:lang}
	Assume $\{\bTheta_t\}$ is obtained from the wrapping of a Langevin diffusion $\{\bX_t\}$ with drift \eqref{eq:lange} given by a strictly positive toroidal density $f$. Assume that the second derivatives of both $f$ and the entries of $V$ are H\"{o}lder continuous, and that $V$ is $2\pi$-periodical. Then $\{\bTheta_t\}$ is the unique toroidal time-reversible diffusion that is ergodic with stationary density $f$ and prescribed $V$.
\end{prop}

The above result roots on \citet{Kent1978}'s characterization of ergodic time-reversible diffusions on manifolds and is particularly useful for constructing OU toroidal analogues. To that aim, consider first the multivariate OU process
\begin{align}
\rd\bX_t=\bA(\bmu-\bX_t)\rd t+\bSigma^\frac{1}{2}\rd \bW_t,\label{eq:MOU}
\end{align}
with $\bmu\in\R^p$, $\bSigma$ a covariance matrix, and $\bA$ such that $\bA^{-1}\bSigma$ is a covariance matrix. This process has sdi equal to $\mathcal{N}\big(\bmu,\frac{1}{2}\bA^{-1}\bSigma\big)$ and, in virtue of \citet{Kent1978}'s characterization, \eqref{eq:MOU} is the unique time-reversible diffusion with Gaussian sdi and constant diffusion coefficient. Therefore, analogues of the OU process in $\T^p$ follow by wrapping Langevin diffusions for toroidal pdfs that are Gaussian analogues. One of them is the vM due to important Gaussian-like characterizations (see Section 2.2.4 of \citet{Jammalamadaka2001}). Nevertheless, the Wrapped Normal (WN) exhibits also important similarities with the Gaussian (\textit{ibid}, Section 2.2.6) and, contrary to the vM, it appears in Gaussian-related limit laws (see Section 4.3.2 of \citet{Mardia1972}). For this reason, as well as for its better tractability, the focus on obtaining an OU-like toroidal process is on the Langevin diffusion associated to a WN density, referred below as the \textit{WN process}. \\

The pdf of a WN in $\T^p$, $\mathrm{WN}\lrp{\bmu,\bSigma}$, is given by $f_{\mathrm{WN}}(\btheta;\bmu,\bSigma):=\sum_{\bk\in\Z^p}\allowbreak\phi_{\bSigma}(\btheta-\bmu+2\bk\pi)$, with $\bmu\in\T^p$, $\bSigma$ a covariance matrix, and $\phi_{\bSigma}$ the pdf of a $\mathcal{N}(\zero,\bSigma)$. Its interpretation is simple: the series recovers the probability mass spread outside $\T^p$, in a periodic fashion, such that $f_{\mathrm{WN}}$ becomes a density in $\T^p$. Set $\mathrm{WN}\lrp{\bmu,\frac{1}{2}\bA^{-1}\bSigma}$ as the sdi to compare with \eqref{eq:MOU}. Then the WN process follows from wrapping \eqref{eq:langcte} for the pdf associated to the previous sdi:
\begin{align}
\rd \bTheta_t=&\,\bA\bigg(\bmu-\bTheta_t-\sum_{\bk\in\Z^p}2\bk\pi w_{\bk}(\bTheta_t)\bigg)\rd t +\bSigma^\frac{1}{2}\rd \bW_t,\label{eq:wnp}\\
w_{\bk}(\btheta):=&\,\frac{ \phi_{\frac{1}{2}\bA^{-1}\bSigma}(\btheta-\bmu+2\bk\pi)}{\sum_{\bm\in\Z^p} \phi_{\frac{1}{2}\bA^{-1}\bSigma}(\btheta-\bmu+2\bm\pi)}.\notag
\end{align}

\begin{figure}[h!]
	\centering
	\includegraphics[width=0.375\textwidth]{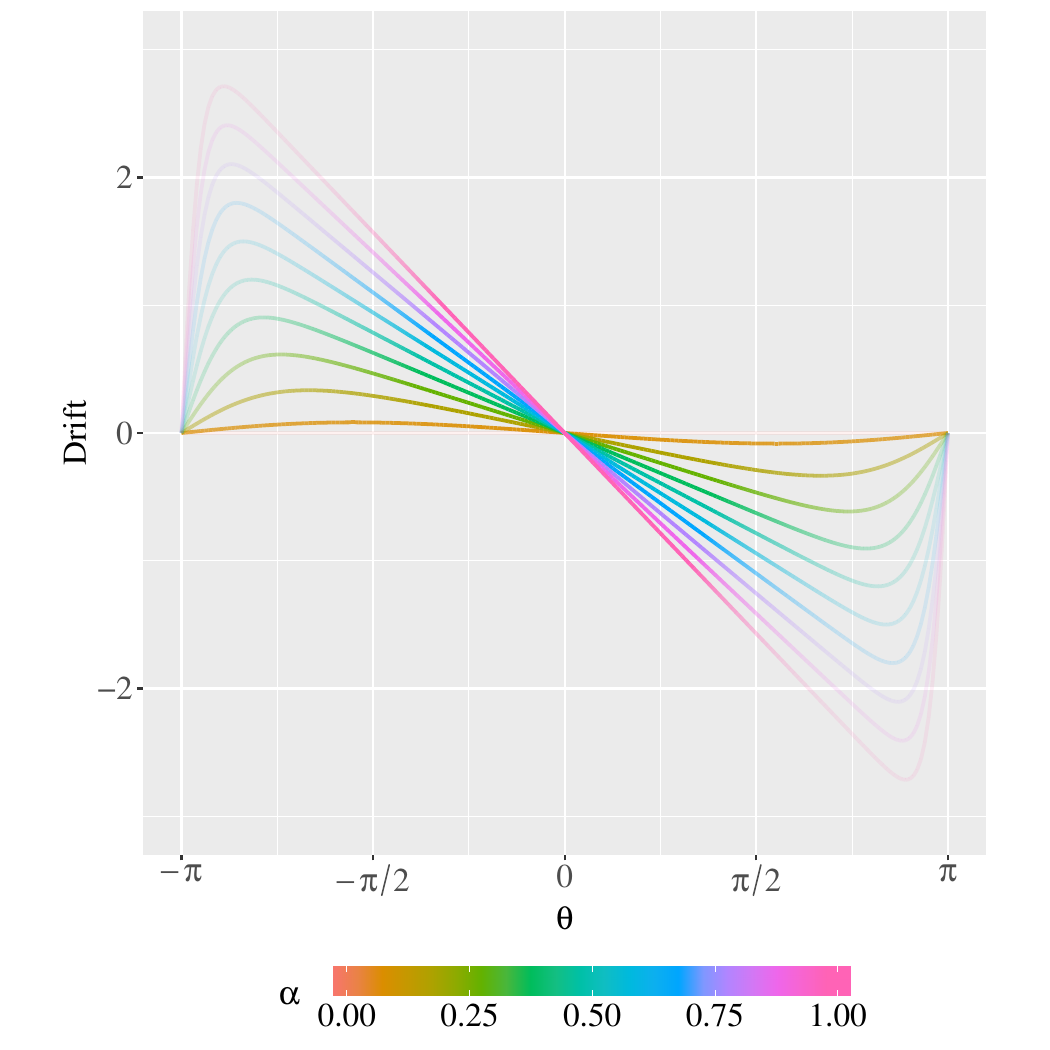}%
	\includegraphics[width=0.375\textwidth]{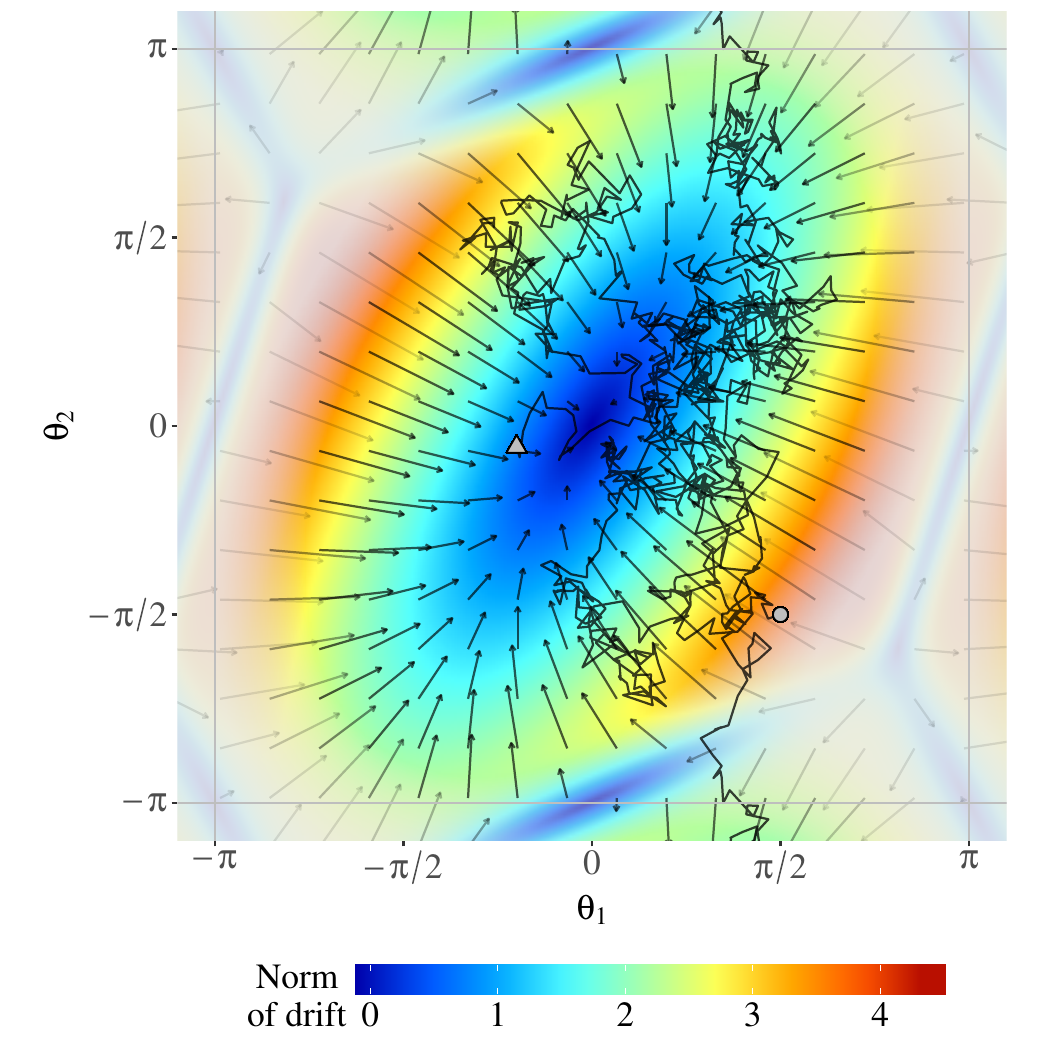}
	\caption{\small Drifts for the WN process for $p=1,2$, with shading proportional to the stationary density. Left: drifts in $p=1$ for $\alpha=i/10$, $i=1,\ldots,10$, $\mu=0$, and $\sigma=1$. Note the periodicity of the drift, its linearity around $\mu$, and how the drift maxima shift towards antipodality as $\alpha$ increases. Right: drift vector field in $p=2$ for $\bA=(1.5, -0.5; -0.5, 1)$, $\bmu=(0,0)$, and $\bSigma=\sigma^2\bI$ with $\sigma=1.5$. The colour gradient represents the intensity of the drift, measured as the norm of the arrows. Note that the stronger drifts are located at the regions with lowest density. A trajectory starting at $(\frac{\pi}{2},-\frac{\pi}{2})$ (round facet) is displayed evolving for $t\in[0,5]$ towards its final point (triangular facet).\label{fig:vecdf}}
\end{figure}

Illustrative drifts of the WN process are shown in Figure \ref{fig:vecdf}. For $p=1$, $\bA=\alpha$ and $\bSigma=\sigma^2$. In this case the WN drift is a smoothed binding of lines with slope close to $-\alpha$ that go through $(\mu,0)$ and such that they are bended to pass through $(\mu\pm\pi,0)$. Hence, the drift behaves almost linearly in a neighbourhood of $\mu$ (equilibrium point, stable) and rapidly decays to pass across $\mu\pm\pi$ (equilibrium point, unstable). The drift maxima vary from $\mu\pm\pi$ (if $\frac{\sigma^2}{2\alpha}\to0$, the sdi is degenerate at $\mu$) to $\mu\pm\frac{\pi}{2}$ (if $\frac{\sigma^2}{2\alpha}\to\infty$, the sdi is uniform and the drift is null). When $p=2$, the vector field of the drift has a characteristic tessellated structure formed by hexagonal-like tiles \textit{anchored} at the points $\bmu+\bk_0\pi$, $\bk_0\in\lrb{-1,0,1}^p\backslash\{\mathbf{0}\}$, where the drift is null. The covariance matrix $\bSigma$ alters the tessellation structure non-trivially by modifying $\{w_{\bk}(\btheta):\bk\in\Z^p\}$. When $\bSigma=\sigma^2\bI$, the larger (respectively, smaller) $\sigma$, the more spread (concentrated) the distribution $\{w_{\bk}(\btheta):\bk\in\Z^p\}$ is, resulting in flat (peaked) drifts with smooth (rough) transitions in the limits defining the tessellation. 

Finally, note that it is easy to parametrize the $2\times2$ drift matrices $\bA$ such that $\bA^{-1}\bSigma$ is a covariance matrix. For $\bSigma=\lrp{
	\sigma_1^2, 0;
	0, \sigma_2^2}$, these are $\bA=\big(
\alpha_1, \frac{\sigma_1}{\sigma_2}\alpha_3;
\frac{\sigma_2}{\sigma_1}\alpha_3, \alpha_2\big)$, with $\alpha_1,\alpha_2>0$ and $\alpha_3^2<\alpha_1\alpha_2$. In particular, the dependence between components is modelled by $\alpha_3$, as is evident from the stationary covariance matrix:
$\tfrac{1}{2}\bA^{-1}\bSigma=\frac{1}{2(\alpha_1\alpha_2-\alpha_3^2)}
\lrp{\alpha_2, \sigma_1^2; -\alpha_3 \sigma_1 \sigma_2; -\alpha_3 \sigma_1 \sigma_2, \alpha_1 \sigma_2^2}$.

\subsection{Estimation for toroidal diffusions}
\label{subsec:est}

The Maximum Likelihood Estimator (MLE) of the parameter $\blambda$ of 
\begin{align}
\rd \bTheta_t=b(\bTheta_t;\blambda)\rd t+\sigma(\bTheta_t;\blambda)\rd \bW_t,\label{eq:param}
\end{align}
when the sample is a discretized trajectory $\{\bTheta_{\Delta i}\}_{i=0}^N$ in the time interval $[0,T]$, $T=N\Delta$, is given by $\hat\blambda_{\mathrm{MLE}}:=\arg\max_{\blambda\in\Lambda} l(\blambda;\{\bTheta_{\Delta i}\}_{i=0}^N)$, where, using the Markovianity of \eqref{eq:param}, the log-likelihood is given by
\begin{align}
l\lrp{\blambda;\{\bTheta_{\Delta i}\}_{i=0}^N}=\log
p(\bTheta_{0};\blambda)+\sum_{i=1}^N\log p_{\Delta}(\bTheta_{\Delta
	i}\,|\,\bTheta_{\Delta (i-1)};\blambda).\label{eq:ll} 
\end{align}
Here $p_{\Delta}(\cdot\,|\,\cdot;\blambda)$ is the tpd of \eqref{eq:param} and the first term in \eqref{eq:ll} is set to the sdi of \eqref{eq:param} if the process is assumed to start in the stationary regime. The MLE can rarely be readily obtained, as usually no explicit expression for the tpd exists. In the following two estimation strategies that rely on an \textit{approximate likelihood function}, where the unknown tpd is replaced by an approximation, are studied. For the sake of brevity, we suppress $\blambda$ in the notation. \\

We note also that the tpd can be computed by solving numerically the PDE \eqref{eq:pde}. This is a computationally expensive task, too demanding for computing the MLE for $p>1$, but useful for obtaining insightful visualizations of the tpd (see Figure \ref{fig:tpds}) and for constructing the accuracy benchmark employed in Section \ref{subsec:sim} to test the more computationally expedient approximate likelihoods. We refer to \citet{Garcia-Portugues:lang} for the details of how to solve this PDE for $p=1,2$.

\subsubsection{Adapted pseudo-likelihoods}

The above PDE solution is too costly for obtaining the MLE. The Euler pseudo-tpd is a cheap computational alternative. The Euler scheme arises as the first order discretization of the process, where the drift and the diffusion coefficient are approximated constantly between 
$\bTheta_{\Delta(i-1)}$ and $\bTheta_{\Delta(i)}$:
\begin{align}
\bTheta_{\Delta i}=\cmod{\bTheta_{\Delta(i-1)}+b(\bTheta_{\Delta(i-1)})\Delta
	+\sqrt{\Delta}\sigma(\bTheta_{\Delta(i-1)})\mathbf{Z}^i},\label{eq:euler}
\end{align}
where $\mathbf{Z}^i\sim\mathcal{N}(\mathbf{0},\bI)$, $i=1,\ldots,N$, and $\cmod{\cdot}=((\cdot+\pi)\mod 2\pi)-\pi$. This
yields the Euler pseudo-tpd 
\[
p^\mathrm{E}_{\Delta}(\btheta\,|\,\bphi)=
f_\mathrm{WN}\lrp{\btheta;\bphi+b(\bphi)\Delta,V(\bphi)\Delta},\quad \btheta,\,\bphi\in\T^p.
\]
Sampling trajectories of \eqref{eq:param} with separation time $\Delta$ can be done by using \eqref{eq:euler} iteratively for a separation time $\Delta/M$, with $M>1$ (thus an $M$ times finer discretized trajectory than the required to sample), and then thinning the trajectory to achieve a separation time $\Delta$. This reduces notably the sampling bias introduced by the Euler scheme. \\

\begin{figure}[h]
	\centering
	\includegraphics[width=\textwidth,clip,trim={0cm 6.5cm 0cm 0cm}]{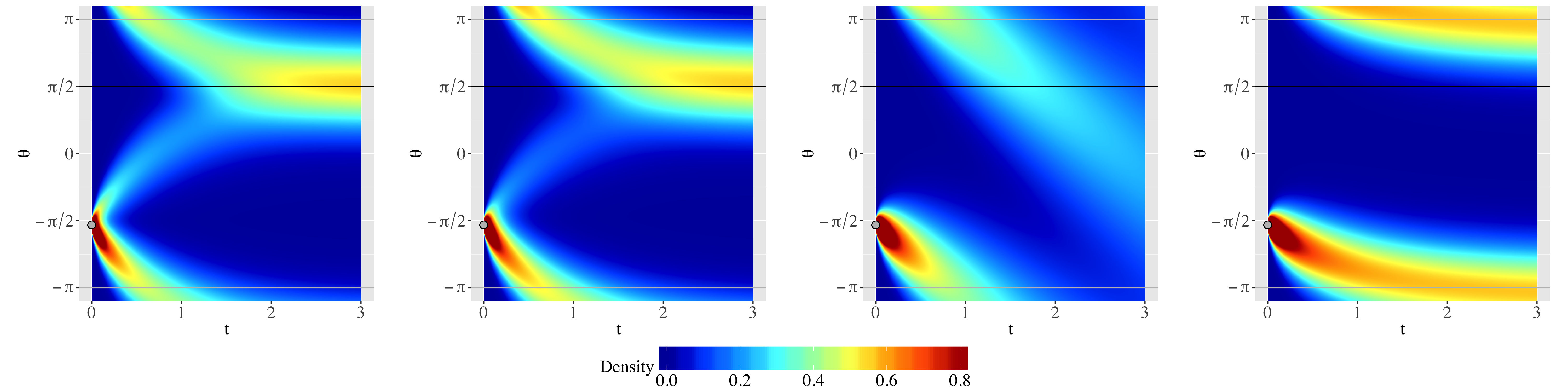}
	\includegraphics[width=\textwidth]{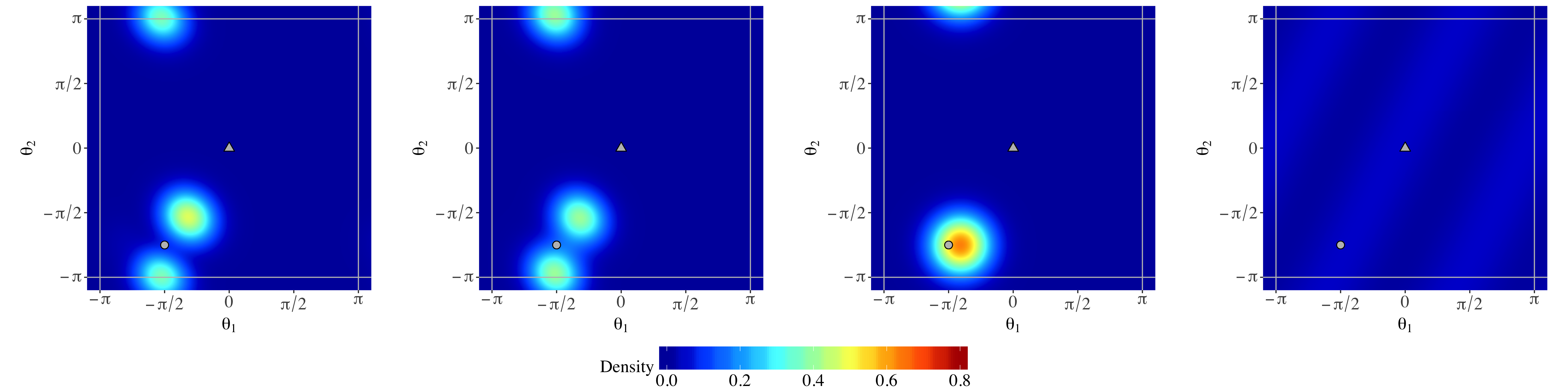}
	\caption{\small From left to right columns: numerical solution of the PDE, WOU tpd approximation, Euler pseudo-tpd, and Shoji--Ozaki pseudo-tpd. Top row: approximations for $p_t(\cdot\,|\,\theta_0)$ with $t\in[0.01,3]$, for the one-dimensional WN process with $\alpha=1$, $\mu=\frac{\pi}{2}$ (horizontal line), $\sigma=1$, and $\theta_0=-\frac{\pi}{2}-0.1$ (round facet). Bottom row: approximations for $p_t(\cdot\,|\,\btheta_0)$ at $t=0.25$, for the two-dimensional WN process with $\bA=(1,0.5;0.5,1)$, $\bmu=(0,0)$ (triangular facet), $\bSigma=\bI$, and $\btheta_0=(-\frac{\pi}{2}, -\frac{3\pi}{4})$ (round facet). Note the \textit{explosion} in the Shoji--Ozaki tpd, a consequence of negative drift eigenvalues at $\btheta_0$. \label{fig:tpds}}
\end{figure}

An improvement on the Euler scheme is the Shoji-- \citep{Shoji1998} scheme. It employs a linear approximation for the drift, $b(\bX_t)\approx b(\bX_s)+\bJ_s(\bX_t-\bX_s)$ for $t\in[s, s + \Delta)$ ($\bJ_s=J(\bX_s)$ denotes the Jacobian of $b$ at $\bX_s$), and approximates the diffusion coefficient constantly. This results in a linear SDE that can be solved explicitly. Wrapping this solution provides the Shoji--Ozaki pseudo-tpd: 
\[
p^\mathrm{SO}_{\Delta}(\btheta\,|\,\bphi)
= f_\mathrm{WN}\lrp{\btheta;E_\Delta(\bphi),V_\Delta(\bphi)},\quad \btheta,\,\bphi\in\T^p, 
\]
where, assuming that $V(\bphi)^{-1}J(\bphi)$ is symmetric (the case for Langevin diffusions), $E_\Delta(\bphi)=\bphi+J(\bphi)^{-1}(\exp\{J(\bphi)\Delta\}-\bI)b(\bphi)$ and $V_\Delta(\bphi)=\frac{1}{2}J(\bphi)^{-1}(\exp\{2J(\bphi)\Delta\}-\bI)V(\bphi)$. If the real parts of the eigenvalues of $J(\bphi)$ are negative, then  $p^\mathrm{SO}_{\Delta}(\btheta\,|\,\bphi)\allowbreak\underset{\Delta\to\infty}{\longrightarrow} f_\mathrm{WN}\lrp{\btheta;\bphi - J(\bphi)b(\bphi),-\frac{1}{2}J(\bphi)^{-1}V(\bphi)}$ (see top row of Figure \ref{fig:tpds}). Otherwise, the pseudo-tpd degenerates into a uniform density when $\Delta\to\infty$, as Euler's tpd always does (see Figure \ref{fig:tpds}). The Euler and Shoji--Ozaki pseudo-likelihoods follow by replacing the tpd by the pseudo-tpds in \eqref{eq:ll}. 

\subsubsection{Transition density approximation for the WN process}

We consider now a specific analytic approximation for the tpd of the WN process aimed to work equally well irrespectively of $\Delta$ (a vital aspect for ETDBN where there is little control over $\Delta$) and to cope with its potential multimodality (the number of potential modes is $2p$). The approximation relies on the connection of the WN process with \eqref{eq:MOU}, whose tpd is $p_t(\cdot\,|\,\bx_s;\bA,\bmu,\bSigma)=\phi_{\bGamma_t}(\cdot-\bmu_t)$, where $\bmu_t :=\bmu + e^{-t\bA }(\bx_s - \bmu)$ and $\bGamma_t:=\int_0^te^{-s\bA}\bSigma e^{-s\bA'}\rd s$. We denote by WOU, standing for Wrapped multivariate OU process, to the wrapping of the process \eqref{eq:MOU}. Assuming that $\bX_s\sim \mathcal{N}\big(\bmu,\frac{1}{2}\bA^{-1}\bSigma\big)$, the conditional density of the WOU process follows from the tpd of \eqref{eq:MOU} as
\begin{align}
p^\mathrm{WOU}_t(\btheta\,|\,\btheta_s;\bA,\bmu,\bSigma) := \sum_{\bm\in\Z^p} f_\mathrm{WN}(\btheta;\bmu^\bm_t,\bGamma_t)w_{\bm}(\btheta_s)
\label{eq:wou}
\end{align}
where $\bmu^\bm_t :=\bmu + e^{-t\bA }(\btheta_s - \bmu + 2\bm\pi)$. The conditional density \eqref{eq:wou} is the wrapping of the tpd of \eqref{eq:MOU} plus a weighting by the sdi of the winding numbers of $\bX_s$, resembling the structure of the WN drift: a weighting of linear drifts by the winding number sdi to achieve periodicity. Albeit \eqref{eq:wou} is \textit{not} the tpd of the WN process, both behave similarly in several key situations, as shown in the next result. Note that sampling from \eqref{eq:wou} is immediate and requires no intermediate steps as in \eqref{eq:euler}: \textit{i}) simulate $\bM$ from the discrete distribution $\Prob{\bM=\bm}=w_{\bm}(\btheta_s)$, $\bm\in\Z^p$; \textit{ii}) sample from a $\mathcal{N}(\bmu_t^\bM,\bGamma_t)$ and wrap the output by $\cmod{\cdot}$. 

\begin{coro}[\citet{Garcia-Portugues:lang}]
	\label{coro:wou}
	{\em Suppose $\bTheta_t$ solves \eqref{eq:wnp} with $\bTheta_0 =
		\btheta_0$ and let $\bTheta^\mathrm{WOU}_t$ be the wrapping of the solution to \eqref{eq:MOU}, where $\bX_0\sim\mathcal{N}(\bmu,\frac{1}{2}\bA^{-1}\bSigma)$ and $\bTheta^\mathrm{WOU}_0 =\btheta_0$. Then $p^\mathrm{WOU}_t$ approximates $p_t$, the true tpd of the WN process, due to the following facts (the dependence on the parameters is omitted):
		\begin{enumerate}[label=\textit{\roman*.}, ref=\textit{\roman*}]
			\item \textit{Point mass}: as $t \rightarrow 0$, $\bTheta_t\rightarrow\btheta_0$ and $\bTheta^\mathrm{WOU}_t\rightarrow\btheta_0$ in probability. 
			\item \textit{Sdi-correct}: as $t \rightarrow \infty$, both $\bTheta_t$ and $\bTheta^\mathrm{WOU}_t$ converge to a $\mathrm{WN}(\bmu,\tfrac{1}{2}\bA^{-1}\bSigma)$ in distribution.
			\item \textit{Time-reversibility}: as $p_t$ does, $p_t^\mathrm{WOU}$ satisfies
			\[
			f_{\mathrm{WN}}(\btheta_0;\bmu,\tfrac{1}{2}\bA^{-1}\bSigma)
			p^\mathrm{WOU}_t(\btheta\,|\,\btheta_0) =
			f_{\mathrm{WN}}(\btheta;\bmu,\tfrac{1}{2}\bA^{-1}\bSigma)
			p^\mathrm{WOU}_t(\btheta_0\,|\,\btheta).
			\]
			\item \textit{High-concentration}: if $\bA^{-1}\bSigma \rightarrow \mathbf{0}$ with $\bSigma$ bounded,
			$\bTheta_t - \bTheta^\mathrm{WOU}_t \rightarrow \mathbf{0}$ in probability, so the
			distributions of $\bTheta_t$ and $\bTheta^\mathrm{WOU}_t$ are similar in the limit.
	\end{enumerate}}
\end{coro}

The tractability of \eqref{eq:wou} degenerates quickly with the dimension, but it can be readily computed for $p=2$ by a series of computational tricks. Specifically, $e^{-t\bA}$ can be obtained in virtue of Corollary 2.4 of \citet{Bernstein1993}: for any $2\times2$ matrix $\bA$, $e^{t\bA}=a(t)\bI + b(t)\bA$ with $a(t):=e^{s(\bA)t}\big(\cosh(q(\bA)t)-s(\bA)\tfrac{\sinh(q(\bA)t)}{q(\bA)}\big)$, $b(t):=e^{s(\bA)t}\tfrac{\sinh(q(\bA)t)}{q(\bA)}$, $s(\bA):=\tfrac{\tr{\bA}}{2}$, and $q(\bA):=\sqrt{\abs{\det(\bA-s\bI)}}$. The fact that $\bA^{-1}\bSigma$ is symmetric plus the previous formula gives
\begin{align*}
\boldsymbol{\Gamma}_t=\frac{1}{2}\bA^{-1}(\bI-\exp\{-2\bA t\})\bSigma=s(t)\frac{1}{2}\bA^{-1}\bSigma + i(t)\bSigma,
\end{align*}
with $s(t)=1-a(-2t)$ and $i(t)=-\tfrac{1}{2}b(-2t)$. This expression gives a neat interpolation between the infinitesimal and stationary covariance matrices, specifically convenient for computing the tpd for several $t$'s. 

\subsection{Empirical performance}
\label{subsec:sim}

The goodness-of-fit of the tpd approximations has a direct influence on the resulting approximate MLEs for the WN process. To measure the closeness of the approximation, we consider the Kullback--Leibler (KL) divergence of the approximation $p^\mathrm{A}_t(\cdot\,|\,\btheta_s)$ ($\mathrm{A}=\mathrm{E},\mathrm{SO},\mathrm{WOU}$) to $p_t(\cdot\,|\,\btheta_s)$ by \textit{weighting} by the sdi $\mathrm{WN}\lrp{\bmu,\frac{1}{2}\bA^{-1}\bSigma}$ the contributions of each initial point $\btheta_s$ to the divergence. Since the PDE solution is obtained for an initial condition of the form $\mathrm{WN}(\btheta_s,\sigma_0^2\bI)$, we consider the same initial condition for the approximations to remove the bias in the comparison: $p^\mathrm{A}_{t,\sigma_0^2}(\bthe\,|\,\btheta_s):=\int_{\T^p}p^\mathrm{A}_t(\bthe\,|\,\bphi)f_\mathrm{WN}(\bphi;\bthe_s,\sigma_0^2)\rd \bphi$. The divergence measure we consider is then
\[
\mathrm{D}^\mathrm{A}_{t,\sigma_0^2}:=\int_{\T^p}\int_{\T^p}p^{\mathrm{PDE}}_{t,\sigma_0^2}(\btheta\,|\,\btheta_s)\log\lrp{\frac{p^{\mathrm{PDE}}_{t,\sigma_0^2}(\btheta\,|\,\btheta_s)}{p^\mathrm{A}_{t,\sigma_0^2}(\btheta\,|\,\btheta_s)}}f_{\mathrm{WN}}\big(\btheta_s; \bmu, \tfrac{1}{2}\bA^{-1}\bSigma\big)\rd\btheta\rd\btheta_s.
\]

\begin{figure}[!h]
	\centering
	\includegraphics[width=0.9\textwidth]{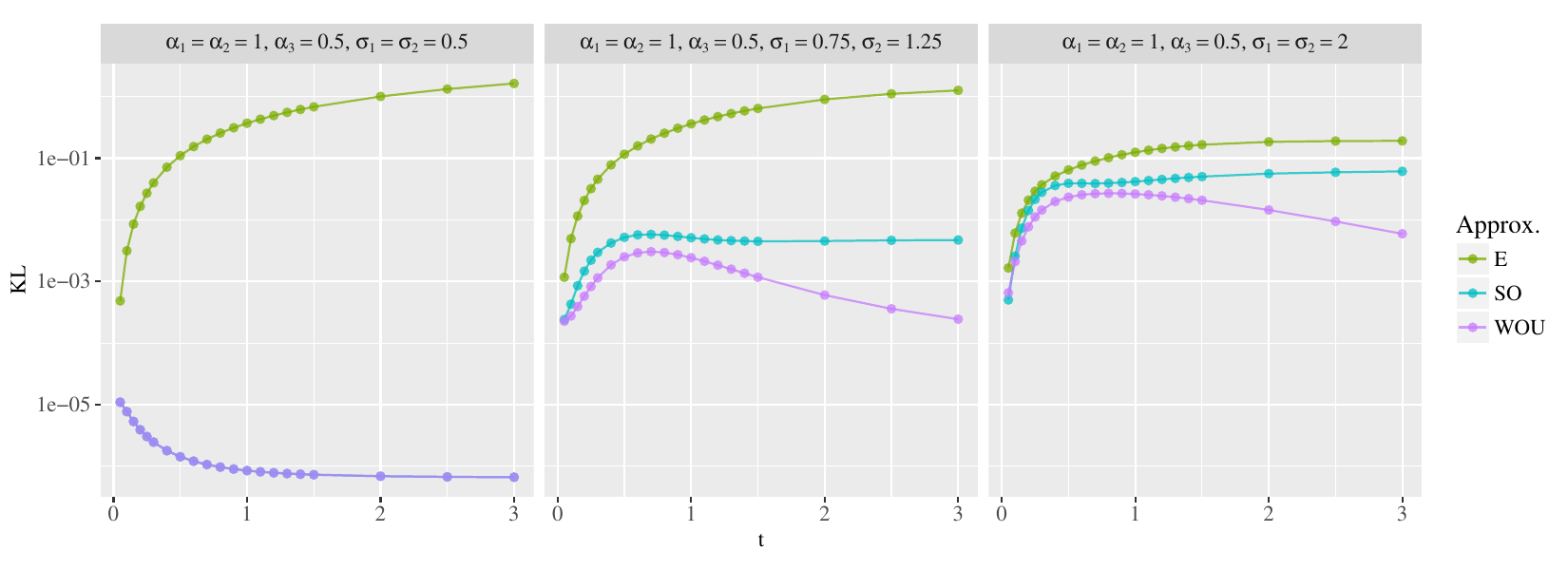}
	\caption{\small $\mathrm{D}^\mathrm{A}_{t,\sigma_0^2}$ curves of the WN process with $p=2$. Note the vertical log-scale. From left to right, panels represent small, moderate, and high diffusivities. The PDE was solved with $M_x=M_y=240$, $M_t=\lceil 1500t\rceil$, and $\sigma_0=0.1$.}
	\label{fig:klwn2d}
\end{figure}

Figure \ref{fig:klwn2d} shows the $\mathrm{D}^\mathrm{A}_{t,\sigma_0^2}$ curves in log-scale for the WN process with $p=2$, under three illustrative drift strengths and diffusivities. As it can be seen, WOU outperforms the other approximations under all scenarios and times. Besides, WOU is the only approximation whose accuracy improves as time increases (above a certain local maximum in the KL divergence), whereas the E and SO pseudo-tpds either deteriorate or stabilize as time increases. The exception is the scenario with low diffusivity where SO is almost equal to WOU (and both are close to the true tpd). E is systematically behind SO in performance. Further empirical results in \citet{Garcia-Portugues:lang} corroborate this pattern. \\

We compare now the efficiency of WOU, SO, and E in estimating the unknown parameters of the WN process in $p=2$ from a sample of $N=250$ points. We consider $\Delta=0.05,0.20,0.50,1.00$ and four representative parameter choices for the WN process. The trajectories are simulated using the E method with time step $0.001$ and then subsampled for given $\Delta$'s. $\bSigma$ is assumed to be known to avoid the inherent unidentifiabilities of $\bA$ and $\bSigma$ when $\Delta$ is large and the tpd converges to the sdi. To summarize the overall performance of the three estimators (E, SO, and WOU) of the $5$-variate parameter $\blambda=(\alpha_1,\alpha_2,\alpha_3,\mu_1,\mu_2)$, a global measure of relative performance is considered. This measure is the componentwise average of Relative Efficiency (RE), where the RE is measured with respect to the best estimator at a given component in terms of Mean Squared Error (MSE). Hence, if $\hat\blambda_j$ ($j\in\{\mathrm{E},\mathrm{SO},\mathrm{WOU}\}$) is the best estimator for all the components of $\blambda$, then $\mathrm{RE}(\hat\blambda_j)=1$. $\mathrm{RE}(\hat\blambda_j)$ is estimated with $1000$ Monte Carlo replicates and $\hat\blambda_j$ is obtained by maximizing the approximate likelihood with a common optimization procedure that employs stationary estimates as starting values. \\

\begin{table}[h]
	\centering
	\small
	\begin{tabular}{c|ccc||ccc}
		& \multicolumn{3}{c||}{$\alpha=1$,\quad $\sigma=1$} & \multicolumn{3}{c}{$\alpha=2$,\quad $\sigma=1$} \\ \midrule
		$\Delta$ &   E   &   SO   &     WOU     &  E  &  SO  &       WOU       \\ \midrule
		$0.05$ & \bf 0.9765 & 0.9244 & 0.8999 & \bf 0.9920 & 0.8452 & 0.8460\\
		$0.20$ & \bf 0.9985 & 0.8214 & 0.8229 & 0.7234 & 0.9978 & \bf 0.9993\\
		$0.50$ & 0.5679 & 0.9868 & \bf 0.9972 & 0.4370 & \bf 1.0000 & 0.9980\\
		$1.00$ & 0.4296 & 0.9872 & \bf 0.9998 & 0.3467 & \bf 1.0000 & 0.9970\\ \midrule
		& \multicolumn{3}{c||}{$\alpha=1$,\quad $\sigma=2$} & \multicolumn{3}{c}{$\alpha=2$,\quad $\sigma=2$} \\ \midrule
		$\Delta$ &   E   &   SO   &     WOU     &  E  &  SO  &       WOU       \\ \midrule
		$0.05$ & 0.9297 & \bf 1.0000 & 0.9422 & \bf 0.9635 & 0.8752 & 0.8793\\
		$0.20$ & 0.8249 & 0.9573 & \bf 0.9916 & 0.6017 & 0.7333 & \bf 1.0000\\
		$0.50$ & 0.6050 & 0.6607 & \bf 1.0000 & 0.3797 & 0.6406 & \bf 1.0000\\
		$1.00$ & 0.5254 & 0.5432 & \bf 1.0000 & 0.2690 & 0.4214 & \bf 1.0000\\ 
	\end{tabular}
	\caption{\small Relative efficiencies for the WN process with $p=2$, $\bmu=\lrp{\tfrac{\pi}{2},-\tfrac{\pi}{2}}$, $\alpha_1=\alpha_2=\alpha$, $\alpha_3=\tfrac{\alpha}{2}$, and $\bSigma=\sigma^2\bI$. Bold font denotes the most efficient estimator for each row and scenario. \label{tab:wn2}}
\end{table}

Table \ref{tab:wn2} gives the REs for E, SO, and WOU in $p=2$. When averaging across scenarios and discretization times, the global ranking of performance is: WOU ($0.9608$), SO ($0.8372$), and E ($0.6607$). E is the best performing method for $\Delta=0.05$ but its relative efficiency quickly decays as $\Delta$ increases. SO and WOU perform similarly for low diffusive scenarios ($\sigma=1$), but for $\sigma=2$ WOU significantly outperforms SO for $\Delta=0.20,050,1.00$, a fact explained by the proneness of the tpd to be multimodal in those situations. The competitive performance of WOU under all scenarios and $\Delta$'s, in addition to its affordable computational cost, places it as the preferred estimation method for the WN process with $p=2$. Similar empirical results hold for $p=1$, see \citet{Garcia-Portugues:lang} for details. \\

As a conclusion, the WN process is seen to be a suitable toroidal diffusion for the needs of ETDBN: OU-like toroidal diffusion, with known sdi, time-reversible, and with tractable inference.

\section{ETDBN: an evolutionary model for protein pairs}
\label{sec:etdbn}

Shortly stated, ETDBN \citep{Golden:evo} is a generative evolutionary model of protein sequence and structure evolution that accounts for evolutionary dependencies due to shared ancestry, dependencies between sequence and structure, and local dependencies between aligned sites. Throughout this section we detail structure, training, and benchmarking of ETDBN.

\subsection{Hidden Markov model structure}

\begin{figure}[!h]
	\centering
	\includegraphics[width=0.9\textwidth, clip, trim=1cm 0.2cm 1cm 6.2cm]{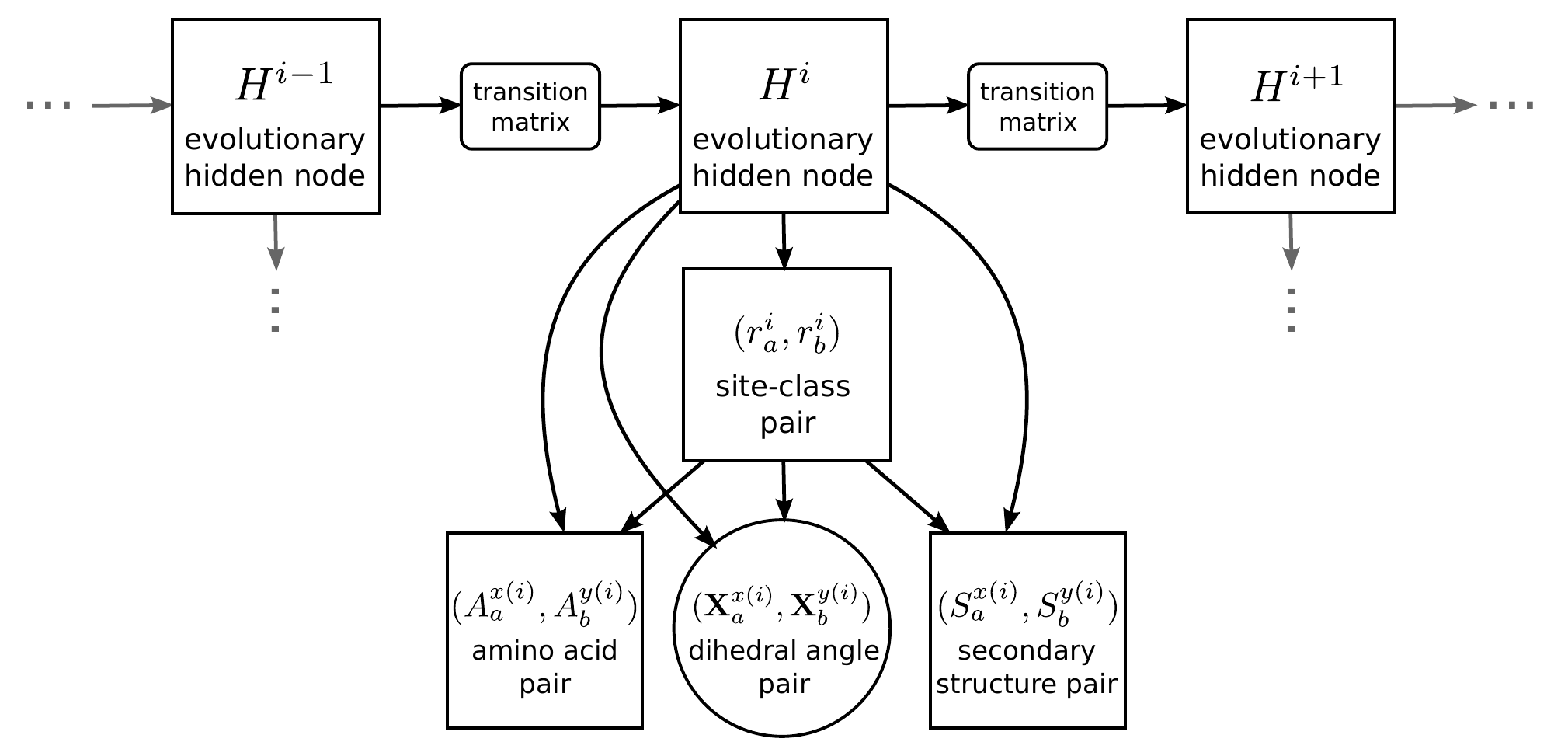}
	\caption{\small HMM architecture of ETDBN. Every hidden node corresponds to a sequence aligned pair of amino acid positions of two homologous proteins $\bP_a$ and $\bP_b$. The horizontal edges between evolutionary hidden nodes encode neighbouring dependencies between aligned amino acid positions. The arrows between the evolutionary hidden nodes and site-class pair nodes encode the conditional independence between the observation pair variables $\big(A_a^{x(i)},A_b^{y(i)}\big)$ (amino acid site pair), $\big(\bX_a^{x(i)},\bX_b^{y(i)}\big)=\big(\big\langle\phi_a^{x(i)},\psi_a^{x(i)}\big\rangle,\big\langle\phi_b^{y(i)},\psi_b^{y(i)}\big\rangle\big)$ (dihedral angle site pair), and $\big(S_a^{x(i)},S_b^{y(i)}\big)$ (secondary structure class site pair). The circles represent continuous variables and the rectangles represent discrete variables.}
	\label{fig:hmm_diagram}%
\end{figure}

ETDBN is a dynamic Bayesian network model for a pair of aligned homologous proteins which can be viewed as a Hidden Markov Model (HMM; see Figure~\ref{fig:hmm_diagram}). Each hidden node corresponds to an aligned site in a sequence alignment and adopts an evolutionary hidden state specifying a distribution over three different observations pairs: a pair of amino acid characters, a pair of dihedral angles, and a pair of secondary structures classifications. A transition probability matrix specifies neighbouring dependencies between adjacent evolutionary states along the alignment. For example, a transition from a hidden state encoding predominantly $\alpha$-helix evolution to a hidden state also encoding $\alpha$-helix evolution is expected to occur more frequently than a transition to a hidden state encoding $\beta$-sheet evolution. \\

The sequence of hidden nodes in the HMM is denoted as $\bH:=(H^{1},H^{2},\ldots,H^{m})$, where $m$ is the length of the sequence alignment $\bM_{ab}:=(M_{ab}^{1},\ldots,M_{ab}^{m})$. Each hidden node $H^{i}$ in the HMM corresponds to a \textit{site observation} pair $\big(\bP_{a}^{x(i)},\bP_{b}^{y(i)}\big)$ at an aligned site $i$ in $\bM_{ab}$ of the two homologous proteins $\bP_a=\big(\bP_a^1,\ldots,\bP_a^{|\bP_a|}\big)$ and $\bP_b=\big(\bP_b^1,\ldots,\bP_b^{|\bP_b|}\big)$. For $i=1,\ldots,m$, $M_{ab}^{i}\in\left\{\left(\substack{x(i)\\y(i)}\right),\left(\substack{x(i)\\-}\right),\left(\substack{-\\y(i)}\right)\right\}$ specifies the homology relationship at position $i$ of the alignment: homologous (no insertions or deletions), deletion with respect to $\bP_a$, and insertion with respect to $\bP_a$, respectively. $x(i)\in\{1,\ldots,|\bP_a|\}$ and $y(i)\in\{1,\ldots,|\bP_b|\}$ specify the indices of the positions in $\bP_a$ and $\bP_b$, respectively, and $|\bP|$ denotes the number of amino acids of protein $\bP$. To simplify the discussion that follows, we treat the sequence alignment $\bM_{ab}$ as given \textit{a priori}, but in practice we extend the HMM to marginalise out an unobserved alignment (see \citet{Golden:evo} for more details). Therefore, we exclude $\bM_{ab}$ in the equations that follow to make our descriptions more concise. \\

ETDBN is parametrised by $h$ hidden states. Thus, every hidden node $H^{i}$ corresponding to an aligned site $i$ can take $h$ possible hidden states and the HMM transition matrix is $h\times h$. Each hidden state specifies a distribution over a \textit{site-class} pair $(r_{a}^{i},r_{b}^{i})$ that is controlled by the evolutionary time $t_{ab}$. A site-class pair consists of two site-classes, $r_{a}^{i}$ and $r_{b}^{i}$, each of the two site-classes taking two integer values, \textit{i.e.} $(r_{a}^{i},r_{b}^{i})\in R:=\{(1,1),(1,2),(2,1),(2,2)\}$. Briefly stated, these site-classes serve for encoding two types of evolution. This is discussed further in Section \ref{subsec:siteclasses}. \\

The state of $H^i$, together with the site-class pair, and the evolutionary time separating proteins $\bP_a$ and $\bP_b$, $t_{ab}$, specify a distribution over three conditionally independent stochastic processes describing each of the three types of site observation pairs: $\big(A_{a}^{x(i)},A_{b}^{y(i)}\big)$, $(\bX_{a}^{x(i)},\bX_{b}^{y(i)}\big)$, and $\big(S_{a}^{x(i)},S_{b}^{y(i)}\big)$. This conditional independence structure allows the likelihood of a site observation pair at an aligned site $i$ to be written as follows:
\begin{align}
p\big(\bP_{a}^{x(i)},\bP_{b}^{y(i)}\,|\,H^{i},r_{a}^{i},r_{b}^{i},t_{ab}\big) := & \,\overbrace{p\big(A_{a}^{x(i)},A_{b}^{y(i)}\,|\,H^{i},r_{a}^{i},r_{b}^{i},t_{ab}\big)}^{\text{amino acid evolution}} \notag\\
& \times \overbrace{p\big(\bX_{a}^{x(i)},\bX_{b}^{y(i)}\,|\,H^{i},r_{a}^{i},r_{b}^{i},t_{ab}\big)}^\text{dihedral angle evolution}\notag\\
& \times \overbrace{p\big(S_{a}^{x(i)},S_{b}^{y(i)}\,|\,H^{i},r_{a}^{i},r_{b}^{i},t_{ab}\big)}^\text{secondary structure evolution},
\label{eq:conditional_independence}
\end{align}
where $p(X)$ denotes either the pdf or the probability mass function of the random variable $X$. The assumption of conditional independence provides computational tractability, allowing us to avoid costly marginalisation when certain combinations of data are missing (e.g. amino acid sequences present, but secondary structures and dihedral angles missing). \\

The amino acid and secondary structure evolution terms in equation \eqref{eq:conditional_independence} are modelled using time-reversible Continuous-Time Markov Chains (CTMC). CTMCs are standard tools for modelling the evolution of discrete states, such as amino acid characters or secondary structure classes. The parameters of the amino acid and secondary structure CTMCs are specified by the hidden state, $H^i$, and evolutionary site-classes $(r^i_a, r^i_b)$. Each CTMC shares a symmetric rate matrix across all hidden states and stationary frequencies specific to each value of $H^i$ and $(r^i_a, r^i_b)$ (see details in \citet{Golden:evo}). The amino acid CTMC contains $20$ states and the secondary structure classes has $3$: Helix (H), Sheet (S), and Coil (C). The joint time-dependent pdfs in \eqref{eq:conditional_independence} are obtained using the pulley principle (Figure \ref{fig:reversibility}) and the transition probabilities of the CTMCs. The dihedral angle evolution is modelled using the bivariate WN process introduced in \eqref{eq:wou} and, using the approximate tpd given in \eqref{eq:wou}, $p\big(\bX_{a}^{x(i)},\bX_{b}^{y(i)}\,|\,H^{i},r_{a}^{i},r_{b}^{i},t_{ab}\big):=p^\textrm{WOU}_{t_{ab}}\big(\bX_{b}^{y(i)}\,|\,\bX_{a}^{x(i)};\bA^\ell,\bmu^\ell,\bSigma^\ell\big)f_\mathrm{WN}\big(\bX_{a}^{x(i)};\bA^\ell,\bmu^\ell,\bSigma^\ell\big)$, where $\ell$ depends on the value of $(H^{i},r_{a}^{i},r_{b}^{i})$. The time-reversibility of $p^\textrm{WOU}_t$ (Corollary \ref{coro:wou}) implies that the roles of $a$ and $b$ are exchangeable in the above expression. \\ \nowidow[3]

Finally, note that the HMM structure and the sequence alignment entail that the joint pdf of a pair of related proteins $\bP_a=(\bA_a,\bX_a,\bS_a)$ and $\bP_b=(\bA_b,\bX_b,\bS_b)$ is given by
\begin{align}
p({\bP}_a,{\bP}_b\,|\,t_{ab})=&\,\sum_{\bH}p({\bP}_{a},{\bP}_{b}\,|\,{\bH},t_{ab})p(\bH)\notag\\
=&\,\sum_{\bH}p\big(\bP_a^{x(1)},\bP_b^{y(1)}\,|\,H^{1},t_{ab}\big)p(H^{1})\notag\\
&\times \prod_{i=2}^{m}p\big(\bP_a^{x(i)},\bP_b^{y(i)}\,|\,H^{i},t_{ab}\big)p(H^{i}\,|\,H^{i-1}),\label{eq:protslikelihood}
\end{align}
where the sum is carried over the $h^m$ possible sequences $\bH$. The factor $p\big(\bP_a^{x(i)},\bP_b^{y(i)}\,|\,H^{i},t_{ab}\big)$ is given in terms of \eqref{eq:conditional_independence} in \eqref{eq:observationlikelihood}. The summation in \eqref{eq:protslikelihood} can be efficiently computed using the HMM forward algorithm, and hidden node sequences can be efficiently sampled from \eqref{eq:protslikelihood} using the Forward Filtering Backward Sampling (FFBS) algorithm.

\subsection{Site-classes: constant evolution and jump events}
\label{subsec:siteclasses}

We now turn to the meaning of the site-class pairs. Two modes of evolution are modelled: \textit{constant evolution} and \textit{jump events}. Constant evolution occurs when the site-class starting in protein $\bP_a$ at aligned site $i$, namely $r_{a}^{i}$, is the same as the site-class ending in protein $\bP_b$ at aligned site $i$, $r_{b}^{i}$, \textit{i.e.} $r_{a}^{i}=r_{b}^{i}$. \\

As already stated, a site-class specifies the parameters of the three conditionally independent stochastic processes describing evolution. A limitation of constant evolution is that the coupling between the three stochastic processes is somewhat weak. This in part stems from the time-reversibility of the stochastic processes -- swapping the order of one of the three observation pairs at a homologous site, e.g. (glycine, proline) instead of (proline, glycine), does not alter the likelihood in equation \eqref{eq:conditional_independence}. Alternatively restated: a ``directional coupling'' of an amino acid interchange does not inform the \textit{direction} of change in dihedral angle or secondary structure. For example, replacing a glycine in an $\alpha$-helix in one protein with a proline at the homologous position in a second protein is expected to break the $\alpha$-helix and to strongly inform the plausible dihedral angle conformations in the second protein. \\

Ideally, we would consider a model in which the underlying site-classes were not fixed over the evolutionary trajectory separating the two proteins, as in the case of constant evolution as described above, but instead were able to ``evolve'' in time. This would allow occasional switches in the underlying site-class at a particular homologous site, which would create a stronger dependency between amino acid, dihedral angle, and secondary structure evolution that captures the directional coupling we desire. To approximate this ``ideal'' model in a computationally efficient manner we introduce the notion of a jump event. A jump event occurs when $r_{a}^{i}\neq r_{b}^{i}$. Whereas constant evolution is intended to capture angular drift (changes in dihedral angles localised to a region of the \textit{Ramachandran plot} -- see Figure \ref{fig:stationarydistributions}), a jump event is intended to create a directional coupling between amino acid and structure evolution, and is also expected to capture angular shift (large changes in dihedral angles, possibly between distant regions of the Ramachandran plot). \\

The hidden state at node $H^{i}$, together with the evolutionary time $t_{ab}$ separating proteins $\bP_a$ and $\bP_b$, provides a distribution over a site-class pair:
\begin{align}
p(r_{a}^{i},r_{b}^{i}\,|\,H^{i},t_{ab})=p(r_{a}^{i}\,|\,H^i, r_{b}^{i},t_{ab})p(r_{b}^i\,|\,H^i),
\label{eq:jump_jointlikelihood}
\end{align}
where we consider
\begin{align*}
p&(r_{a}^{i}\,|\,H^{i}, r_{b}^{i},t_{ab}):=\begin{cases}
e^{-\gamma_{H^{i}}t_{ab}}+\pi_{H^{i},r_{b}^{i}}(1-e^{-\gamma_{H^{i}}t_{ab}}), & \text{if }r_{a}^{i}=r_{b}^{i},\\
\pi_{H^{i},r_{b}^{i}}(1-e^{-\gamma_{H^{i}}t_{ab}}), & \text{if }r_{a}^{i}\neq r_{b}^{i},\\
\end{cases}
\end{align*}
with $p(r_{a}^{i}\,|\,H^{i}):=\pi_{H^{i},r_{a}^{i}}$ and $p(r_{b}^{i}\,|\,H^{i}):=\pi_{H^{i},r_{b}^{i}}$. Both $\pi_{H^{i},r_{a}^{i}}$ and $\pi_{H^{i},r_{b}^{i}}$ are model parameters specifying the stationary probability of starting starting in site-class $r_{a}^{i}$ and $r_{b}^{i}$, respectively, and therefore are not conditional on the evolutionary time $t_{ab}$. The parameter $\gamma_{H^{i}} > 0$ is specific to the hidden state $H^i$ and determines its associated jump rate. The site-class jump probabilities have been chosen so that time-reversibility holds; in other words it happens that: $p(r_{a}^{i}\,|\,H^i, r_{b}^{i},t_{ab})p(r_{b}^{i}\,|\,H^i)=p(r_{b}^{i}\,|\,H^i,r_{a}^{i},t_{ab})p(r_{a}^{i}\,|\,H^i)$. \\

The hidden state at node $H^i$, together with a site-class pair $(r_{a}^{i},r_{b}^{i})$ and the evolutionary time $t_{ab}$, specifies the likelihood over site observation pairs:
\begin{align}
p\big(\bP_a^{x(i)},&\,\bP_b^{y(i)}\,|\,H^{i},r_{a}^{i},r_{b}^{i},t_{ab}\big)\notag\\
&:=\begin{cases}
p\big(\bP_a^{x(i)},\bP_b^{y(i)}\,|\,H^{i},r_{c}^{i},t_{ab}\big), & \text{if }r_{a}^{i}=r_{b}^{i}=r_{c}^i,\\
p\big(\bP_a^{x(i)}\,|\,H^{i},r_{a}^{i}\big)p\big(\bP_b^{y(i)}\,|\,H^{i},r_{b}^{i}\big), & \text{if }r_{a}^{i}\neq r_{b}^{i}.
\end{cases}
\label{eq:likelihood}
\end{align}
Constant evolution is considered \textit{constant} because each observation type at an aligned site $i$ is drawn from the same stochastic process given by $H^i$ and $r_{c}^{i}$. Note that the strength of the evolutionary dependency within an observation pair depends on $t_{ab}$. \\

In the case of a jump event, the evolutionary processes are, after the evolutionary jump, restarted independently in the sdi of the new site-class. Thus the site observations $\bP_a^{x(i)}$ and $\bP_b^{y(i)}$ are assumed to be drawn from the sdis of two separate stochastic processes corresponding to site-classes $r_{a}^{i}$ and $r_{b}^{i}$, respectively. This implies that, conditional on a jump, the likelihood of the observations is no longer dependent on $t_{ab}$. A jump event can therefore best be thought of as an abstraction that captures the end-points of the evolutionary process, but ignores the potential evolutionary trajectory linking the two site observations. The advantage of abstracting the evolutionary trajectory is that there is no need to perform a computationally expensive integration of all possible trajectories. The likelihood of an observation pair is now simply
\begin{align}
p\big(\bP_a^{x(i)},&\,\bP_b^{y(i)}\,|\,H^{i},t_{ab}\big)\notag\\
&=\sum_{(r_{a}^{i},r_{b}^{i})\in{R}} p\big(\bP_a^{x(i)},\bP_a^{y(i)}\,|\,H^{i},r_{a}^{i},r_{b}^{i},t_{ab}\big) p(r_{a}^{i},r_{b}^{i}\,|\,H^{i},t_{ab}),\label{eq:observationlikelihood}
\end{align}
where \eqref{eq:likelihood} and \eqref{eq:jump_jointlikelihood} provide the terms in \eqref{eq:observationlikelihood}.

\subsection{Model training}
\label{subsec:training}

\subsubsection{Training and test datasets}

In order to train and evaluate ETDBN, a training dataset of $1,200$ protein pairs ($2,400$ proteins; $417,870$ site observation pairs) and a test dataset of $38$ protein pairs ($76$ proteins; $14,125$ site observation pairs) were assembled from $1,032$ protein families in the HOMSTRAD database \citep{mizuguchi1998homstrad} -- a database of homologous protein structures. Dihedral angles were computed from the PDB coordinates of each protein structure using the \texttt{BioPython.PDB} package \citep{hamelryck2003pdb}. The secondary structure at each amino acid position in every protein was annotated using the \texttt{DSSP} software \citep{touw2015series}.

\subsubsection{Model training}

Stochastic Expectation-Maximization (StEM, \citet{Gilks1996}) was used to train the model. StEM is a stochastic version of the well known Expectation-Maximization iterative algorithm \citep{Gilks1996}, which is commonly used for fitting the parameters of HMMs. Its distinguishing feature is that the E-step consists of filling in the values of the latent variables using sampling. StEM is attractive due to its computational efficiency and its tendency to avoid getting stuck in local minima \citep{Gilks1996}. \\

We describe in what follows the E- and M- steps. To that end, let us denote by $\Psi^{(k)}$ the model parameters at the $k$-th iteration. FFBS was used in the E-step to sample hidden node sequences $\bH$ and site-classes ($\br_{a},\br_{b}$). The Metropolis--Hastings algorithm was used to sample $t_{ab}$ for each protein pair in the training dataset conditional on the observations and parameters $\Psi^{(k)}$ (see \eqref{eq:posteriortime} and the description that follows for details). In other words, at iteration $k$ for each pair of aligned observation sequences $\bP_a$ and $\bP_b$ samples were drawn from the following joint distribution:
\begin{align*}
\big(\bH,\br_{a},\br_{b},t_{ab}\big)^{(k)}\sim p\big(\bH,\br_{a},\br_{b},t_{ab}\,|\,\bP_a,\bP_b,\Psi^{(k)}\big).
\end{align*}
In the M-step the samples from the previous E-step were used to update the hidden node parameters ($\Psi^{(k+1)}$) using Efficient Sufficient Statistics (ESSs). For example, the MLE of the transition probability matrix at a particular step can be obtained by calculating the proportion of hidden state transitions within the sampled hidden sequences. Where ESSs were not used, the COBYLA optimization algorithm \citep{powell1994cobyla} in the \texttt{NLOpt} library \citep{johnson2014nlopt} was used to update the parameters.

\subsubsection{Model selection}

Models with hidden states ranging from $8$ to $112$ were trained until convergence for varying numbers of repetitions ($2$ to $4$) using different initial random number seeds. The highest log-likelihood model of each repetition was selected for downstream analysis. Following that, marginal likelihoods $p(\text{data}\,|\,\text{fitted model})$ and corresponding Bayesian Information Criterion (BIC) scores were computed under each model by fixing the alignments to the respective HOMSTRAD alignments. The alignments were fixed \textit{a priori} to make computation of the marginal likelihoods computationally tractable. The 64 hidden state model was selected as the best model, as it presented the lowest BIC. Additionally, predictive accuracies under a homology modelling scenario, $p(\bX_b\,|\,\bA_a,\bA_b,\bX_a,\text{fitted model})$, were calculated for each of the $38$ protein pairs in the test dataset for each of the models. The chosen $64$ hidden state model had predictive accuracies comparable to the model with the highest predictive accuracies (a model with $32$ hidden states). Further details about the model selection can be found in \citet{Golden:evo}. \\

\subsection{Benchmarks}
\label{subsec:empirical}

In this section we perform a series of benchmarks that examine the performance of ETDBN. First, we test how well the model reproduces the empirical distributions of dihedral angles. Next we compare how adding increasingly informative conditioning observations affects uncertainty in the estimates of evolutionary times and improves levels of accuracy in the prediction of dihedral angles. These analyses are facilitated by the conditional independence structure in \eqref{eq:conditional_independence}, which enables computationally efficient posterior inference under different combinations of observed or missing~data. 

\subsubsection{Dihedral angles distribution}

\begin{figure}[!h]
	\centering
	\includegraphics[width=0.75\textwidth]{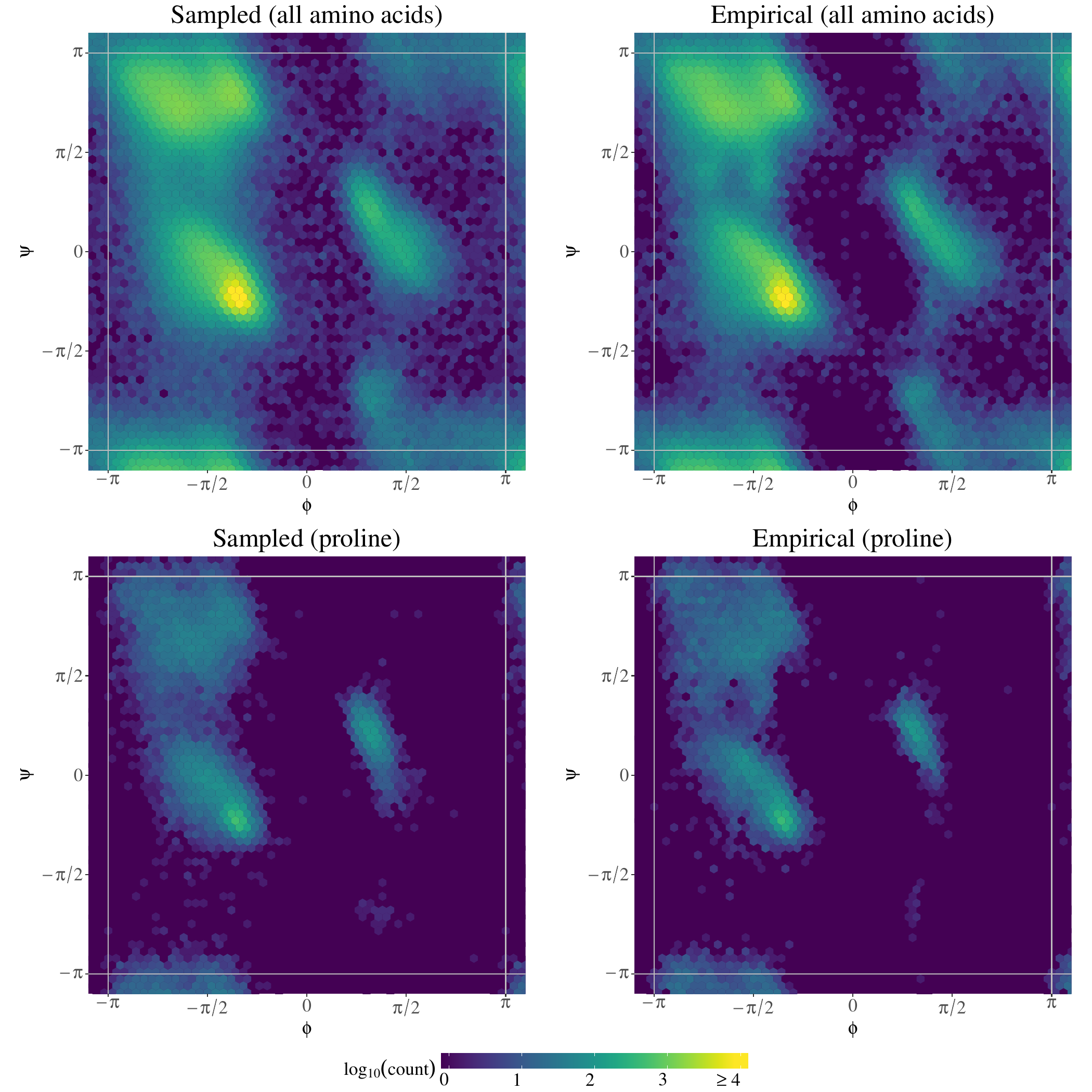}
	\caption{\small Ramachandran plots depicting histograms of the model-sampled dihedral angles (left column) and the dihedral angles in the training dataset (right column). The top row shows the distributions for all amino acids ($417,870$) and the bottom for proline only ($19,324$). The number of angles in the left and right columns is the same.}
	\label{fig:stationarydistributions}%
\end{figure}

The generative nature of ETDBN allows the model to be easily interrogated. In this benchmark, we compare the histograms of dihedral angle pairs present in real data with the dihedrals sampled from ETDBN. Figure~\ref{fig:stationarydistributions} presents this comparison for the $417,870$ dihedral angle pairs present in the training dataset and $19,324$ dihedral angle pairs associated to proline. These distributions are shown in the Ramachandran plots and are a useful tool for visualising the conformational possibilities associated with different amino acids. There is a close correspondence between dihedral angles sampled under the model (Figure~\ref{fig:stationarydistributions}, left) and the empirical distributions (Figure~\ref{fig:stationarydistributions}, right). This serves only as a partial validation of ETDBN, since it was expected that the empirical dihedral angle distributions be well-modelled, given that ETDBN is effectively a mixture model with a large number of mixture components.

\subsubsection{Posterior inference: evolutionary times}

The posterior distribution of $t_{ab}$, conditional on the observed dihedral angles $\bX_a$ and $\bX_b$, is given by Bayes' theorem as
\begin{align}
\label{eq:posteriortime}
p(t_{ab}\,|\,\bX_{a},\bX_{b}) \propto &\, p(\bX_{a},\bX_{b}\,|\,t_{ab})p(t_{ab}) \notag\\
=&\,p(t_{ab})\sum_{\bH}p\big(\bX_{a}^{x(1)},\bX_{b}^{y(1)}\,|\,H^{1},t_{ab}\big)p(H^{1})\notag \\
& \times \prod_{i=2}^mp\big(\bX_{a}^{x(i)},\bX_{b}^{y(i)}\,|\,H^{i},t_{ab}\big)p(H^{i}\,|\,H^{i-1}).
\end{align}
The prior distribution over $p(t_{ab})$ is given by an exponential distribution with $\lambda=\frac{1}{10}$, chosen for its biological plausibility. The Metropolis--Hastings algorithm can be used to sample the posterior distribution, $p(t_{ab}\,|\,\bX_{a},\bX_{b})$, using the associated prior and likelihood in \eqref{eq:posteriortime} (obtained by the HMM structure as in \eqref{eq:protslikelihood}), which can be efficiently computed using the HMM forward algorithm. Analogous expressions hold for $p(t_{ab}\,|\,\bA_{a},\bA_{b})$ and  $p(t_{ab}\,|\,\bA_{a},\bA_{b},\bX_{a},\bX_{b})$. 

The left half of Figure~\ref{fig:benchmarks} depicts the boxplots of posterior standard deviations of the evolutionary times inferred for the proteins in the testing dataset under three different conditions (sequence only, angles only, and both sequence and angles). A one-sided Wilcoxon signed-rank test was used to test whether the standard deviations were significantly smaller when using both types of observations, compared to using only sequence or dihedral angle information. As might be expected, the standard deviations of the posterior evolutionary times were significantly smaller when both the sequence and dihedral angle observations were used for posterior inference, compared to using only sequences ($p\text{-value}=0.00014$) or only dihedral angles ($p\text{-value}=0.00002$). Hence, the model's accuracy on the posterior of $t_{ab}$ improves as more conditioning information is considered. Finally, note also that the sequence alone provides sharper posteriors than the dihedrals alone, a fact likely explained by the direct relation between $t_{ab}$ and $(\bA_a,\bA_b)$.

\subsubsection{Posterior inference: structure}

ETDBN can be used to sample (\textit{i.e.} predict) the dihedral angles of a protein $b$, $\bX_b$, from $\bA_b$, $\bA_a$, $\bX_a$, $\bS_b$, $\bS_a$, or any combination of them. We describe in the next three paragraphs the procedure for achieving so. \\

Conditional on having observed $(\bA_{a},\bA_{b},\bX_{a})$, the sequence of missing dihedral angles $\bX_b$ can be sampled using the following procedure. Firstly, similar to \eqref{eq:posteriortime}, the likelihood $p(t_{ab}\,|\,\bA_{a},\bA_{b},\bX_{a})$ and the prior $p(t_{ab})$ are used to sample a single evolutionary time, ${t_{ab}}^{(k)}$, using  Metropolis--Hastings:
\begin{align}
\label{eq:sampling1}
{t_{ab}}^{(k)}\sim p(t_{ab}\,|\,\bA_{a},\bA_{b},\bX_{a})p(t_{ab}).
\end{align}
When only a single observation is present for a particular observation type (e.g. only $\bX_{a}$ in $p(t_{ab}\,|\,\bA_{a},\bA_{b},\bX_{a})$), it is assumed that the observation is drawn from the sdi of the corresponding stochastic process. Conditional on the observations and newly sampled evolutionary time, $t_{ab}^{(k)}$, the forward probabilities already calculated in \eqref{eq:sampling1} can be used to perform FFBS algorithm to draw a sequence of hidden states, $\bH^{(k)}\sim p\big(\bH\,|\,\bA_{a},\bA_{b},\bX_{a},t_{ab}^{(k)}\big)$. \\

For each hidden state ${H^i}^{(k)}$ in the newly sampled hidden sequence $\bH^{(k)}$, a site-class pair $(r_a^i,r_b^i)^{(k)}$ taking on one of four possible values is sampled with probability proportional (recall an application of Bayes' theorem in the first factor) to $p\big(A_{a}^{x(i)},A_{b}^{y(i)},\bX_{a}^{x(i)}\,|\,{H^i}^{(k)},{r_{a}^{i}},{r_{b}^{i}},{t_{ab}}^{(k)}\big)p\big({r_{a}^{i}},{r_{b}^{i}}\,|\,{H^i}^{(k)},{t_{ab}}^{(k)}\big)$.

\begin{figure}[!h]
	\centering
	\includegraphics[width=0.75\textwidth]{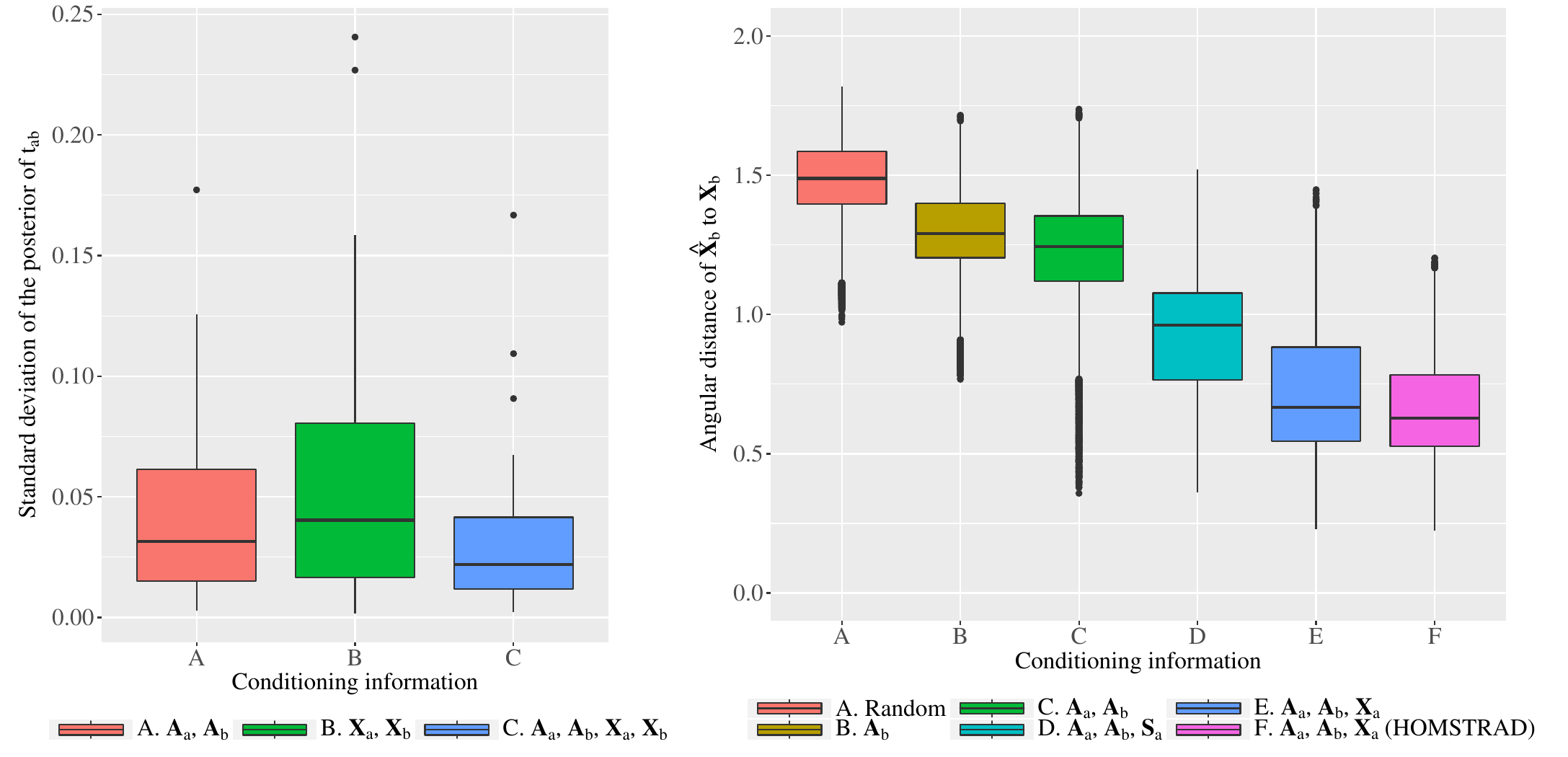}%
	\caption{\small 
		Left: box plots depicting the distribution of posterior standard deviations for the protein pairs in the test dataset under three different scenarios. The mean standard deviations were $0.43$, $0.58$, and $0.33$ for A, B, and C, respectively. Right: comparison of mean predictive accuracy (measured using angular distance, lower is better) for the $38$ protein pairs in the test dataset, giving a representative view of predictive accuracy under six different combinations of observations.}
	\label{fig:benchmarks}
\end{figure}

If ${r_{a}^{i}}^{(k)} \neq {r_{b}^{i}}^{(k)}$, a jump is implied and a dihedral angle pair ${\bX^{y(i)}_{b}}^{(k)}$, corresponding to aligned site $i$, is drawn from the sdi of the diffusion whose parameters are specified by ${r_{b}^{i}}^{(k)}$: ${\bX^{y(i)}_{b}}^{(k)}\sim p\big(\bX_{b}^{y(i)}\,|\,{H^i}^{(k)},{r_{b}^{i}}^{(k)}\big)$. If ${r_{a}^{i}}^{(k)} = {r_{b}^{i}}^{(k)}$, constant evolution is implied and a dihedral angle pair corresponding to aligned site $i$ is drawn from the diffusion whose parameters are specified by ${r_{a}^{i}}^{(k)} = {r_{b}^{i}}^{(k)}$ using the sampling procedure obtained from \eqref{eq:wou}: ${\bX^{y(i)}_{b}}^{(k)}\sim p\big(\bX_{b}^{y(i)}\,|\,{H^i}^{(k)},{r_{b}^{i}}^{(k)},\bX_{a}^{x(i)},{t_{ab}}^{(k)}\big)$. Doing so produces a sequence of dihedral angles, ${\bX_{b}}^{(k)}$, sampled from the correct posterior distribution. \\

For each of the $38$ protein pairs $(\bP_a,\bP_b)$, the dihedral angles of $\bP_b$ in each pair were treated as missing, and $5000$ sequences of $\bX_b$ were sampled under six different combinations of observations for each. Following that, predictive accuracy was measured using the average angular distance between the sampled ($\hat\bX_b$) and known ($\bX_b$) dihedral angles. Figure~\ref{fig:benchmarks} shows the results. In \ref{fig:benchmarks}A, no data was used for prediction, hence the predicted angles correspond to random draws from the model. In \ref{fig:benchmarks}B--\ref{fig:benchmarks}E, the following observations are used, respectively: $\bA_b$, $(\bA_b,\bA_a)$, $(\bA_a,\bA_b,\bS_a)$, and $(\bA_a,\bA_b,\bX_a)$. Finally, in \ref{fig:benchmarks}F the same combination of observations was used as in \ref{fig:benchmarks}E, but the sequence alignment was treated as known \textit{a priori} instead of marginalised. These results show that increasingly informative observations lead to better predictive accuracy, which is consistent with what we expect. \\ \nowidow[3]

\begin{figure}[!h]
	\centering
	\includegraphics[width=0.9\textwidth]{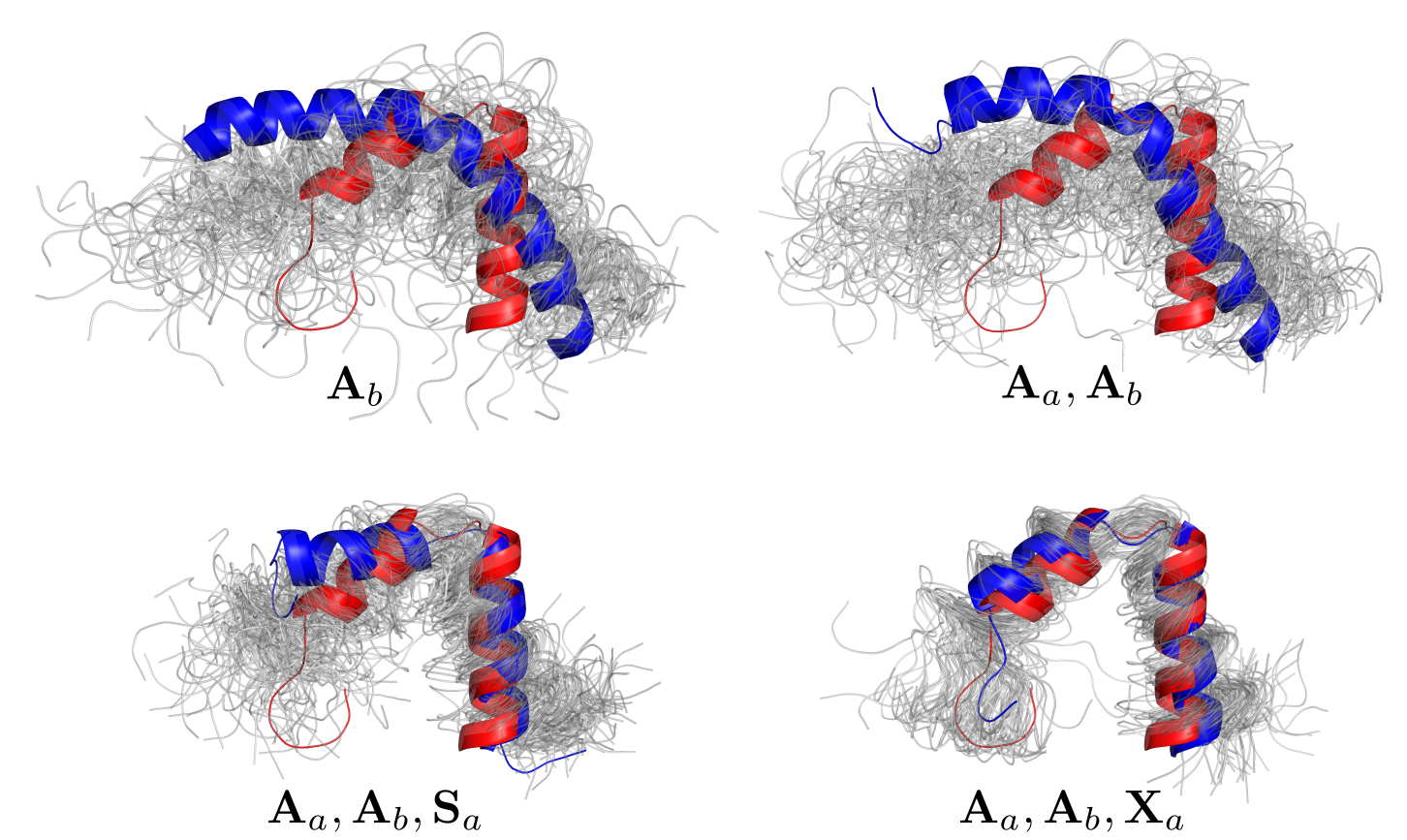}
	\caption{\small Samples from ETDBN of a $35$ amino acid fragment of an annexin protein under increasingly informative conditioning observations (from left to right and up to down). In red is the native protein structure (protein $b$, corresponding to positions $100$--$135$ of PDB 1ala) and in grey is $100$ samples from the model corresponding to the same region. In blue is the centroid structure (sample with the lowest average RMSD to all other samples) based on $2000$ samples from the model. The homologous protein used for the purposes of prediction (protein $a$) was PDB 1ann.}
	\label{fig:prediction}%
\end{figure}

In the right half of Figure~\ref{fig:prediction} we provide a detailed graphical example of a single pair of homologous annexin proteins (PDB 1ala and PDB 1ann) for which sampling was performed under four different conditions. PDB 1ala and PDB 1ann are moderately diverged, having an amino acid sequence identity of 56\%. Whilst the full protein (PDB 1ann) was sampled, only a $35$ amino acid fragment is depicted. The reason is that ETDBN is considered a local model and is not designed to capture the global properties of protein structure. For example, proteins have a strong tendency to be globular (compact) in nature -- a global, coarse-grained property not enforced by our model. Therefore ETDBN does not constitute a complete homology modelling method in itself. Rather, it can be used as a building block (much like fragment libraries model local structure \citep{rohl2004protein}) in protein structure prediction and homology modelling methods. Fine-grained distributions, such as ETDBN, can be combined with with coarse-grained distributions (that capture the globular nature of proteins, for example) in a statistically principled manner using a method known as the \textit{reference ratio method} \citep{hamelryck2010potentials, frellsen2012towards}, an extension out of the scope of this~work. 

Despite these caveats, it is clear once again in Figure~\ref{fig:prediction} that introducing increasingly informative sequence and structural observations lead to better predictions. As more informative data is added the distribution of samples tends to concentrate around the native structure. Increasingly informative conditioning observations also improves the prediction of the secondary structure elements, as can be seen by comparing the patterns of cartoon helices and coils in the centroid and reference structures.

\section{Case study: detection of a novel evolutionary motif}
\label{sec:casestudy}

A benefit of ETDBN is that the $64$ evolutionary hidden states learned during the training phase are interpretable. We give an example of a hidden state encoding a jump event detected in a number of protein pairs in our test and training datasets, suggesting that this hidden state encodes an \textit{evolutionary motif} -- a common pattern of sequence-structure evolution. \\

\begin{figure}[!h]
	\centering
	\includegraphics[width=0.9\textwidth,clip,trim={1.5cm 0.5cm 2cm 0cm}]{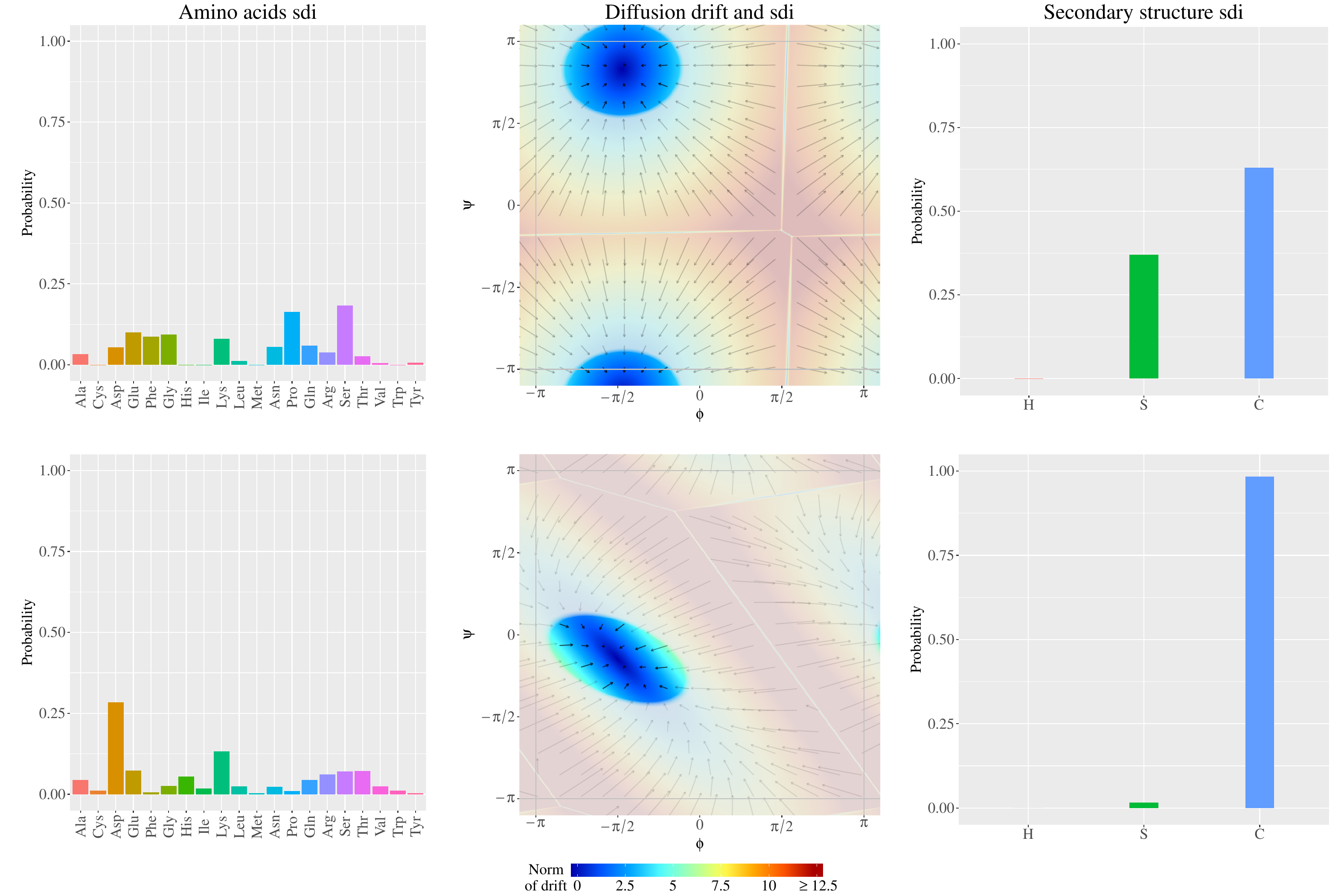}
	\caption{\small Depiction of EHS13 and its associated evolutionary structural motif. This hidden state was sampled at $2.59\%$ of sites (the average was $1.56\%$). The equilibrium frequencies of $r_1$ and $r_2$ were $\pi_1=0.181$ and $\pi_2=0.819$, respectively, and the jump rate was $\gamma=1.32$. The first and second rows depict the parameters encoded by the two site-classes $r_1$ and $r_2$, respectively. The first and third columns show the estimated sdis for the amino acid and secondary structure stochastic processes, respectively. The second column depicts the drift vector fields of the WN processes. The colour gradient represents the intensity of the drift, measured as the norm of the arrows, and with shading proportional to the stationary density.}
	\label{fig:hiddenstate13}%
\end{figure}

Evolutionary Hidden State 13 (Figure~\ref{fig:hiddenstate13}), referred in the sequel as EHS13, was selected from the $64$ hidden states due to its encoding of a jump event with a significant angular shift (a large change in dihedral angle). A notable feature of this EHS13 is that the dihedral angles corresponding to site-classes $r_1$ and $r_2$, respectively, are associated with moderately different amino acid distributions. This is expected to be informative of a specific directional transition in dihedral angles corresponding to a particular directional exchange between amino acids. \\

We performed a search of protein pairs containing the jump represented by ESH13. Posterior inference was performed conditioned on the amino acid sequence and dihedral angles, $(\bA_a,\bA_b,\bX_a,\bX_b)$, of $238$ protein pairs. This was done to identify aligned sites encoding jump events corresponding to potential evolutionary motifs. Aligned sites pertaining to a single hidden state and with evidence of a jump event ($r_a^{i}\neq r_b^{i}$) at posterior probability $>0.99$ were identified, that is, the $i$'s such that $p(H^i,r_a^{i}\neq r_b^{i}\,|\,\bA_a,\bA_b,\bX_a,\bX_b)>0.99$. \\

\begin{figure}[!h]
	\centering
	\includegraphics[width=0.9\textwidth]{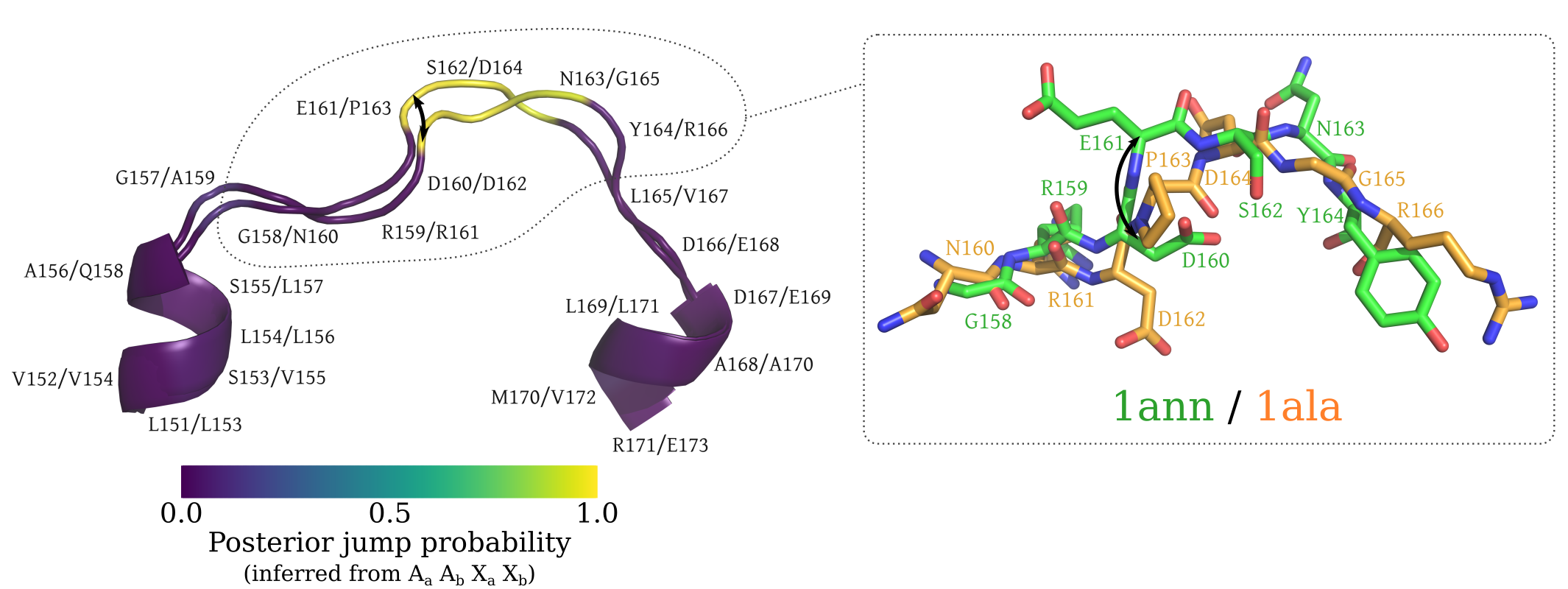}
	\caption{\small Depiction of two homologous annexin proteins, PDB 1ann and 1ala, superimposed. Left: cartoon representation of the two proteins corresponding to regions L151--R171 and L153--E173, respectively, with posterior jump probabilities at each position overlaid. Right: ball-and-stick representation giving atomic detail for a seven amino acid region (G158--Y164 and N160--R166, respectively). The exchange between a glutamate (E161 in 1ann) and a proline (P163 in 1ala) is associated with a large change in dihedral angle as indicated by the curved arrows.}
	\label{fig:jumps}%
\end{figure}

Twelve aligned sites in eleven different protein pairs corresponding to $H^i=13$ (EHS13) were identified. A homologous pair of annexin proteins, PDB 1ann (from \textit{Bos taurus}) and PDB 1ala (from \textit{Gallus gallus}), was selected out of the eleven pairs with a jump corresponding to EHS13 (Figure~\ref{fig:jumps}) at aligned site E161/P163. Most aligned sites in the annexin pair had low posterior jump probabilities ($\approx0.0$), with the exception of three successive aligned sites starting with the aligned site of interest, E161/P163, which had high posterior jump probabilities ($\approx1.0$). Site-classes $\br_1$ and $\br_2$ indicated that an exchange between a proline and a glutamate at positions 161 and 163, respectively, is informative of the dihedral angle change specified by EHS13. \\

It is unknown whether this apparent evolutionary motif has functional consequences, however, the associated local structural and sequence changes between 1ann and 1ala appear significant. The three successive aligned sites with high posterior jump probabilities have dihedral transitions with large angular distances of $1.97$, $1.20$, and $1.96$, respectively, in comparison to the mean angular distance between the dihedral angles of 1ann and 1ala (mean $0.288$ and $90\%$-confidence interval for the mean $(0.034, 0.806)$). All three positions involve an exchange of amino acids. Glutamate to proline (E161/P163) in the first, serine to aspartic acid (S162/D164) in the second, and asparagine to glycine (N163/G165) in the third. The involvement of proline at the first of the three positions in a large conformational shift is unsurprising as proline adopts distinctive dihedral conformations compared to the other amino acids. Likewise, the involvement of a glycine at the third position is also unsurprising given that it is the smallest and most flexible amino acid in terms of the dihedral angle conformations it can adopt. The transition to two such flexible amino acids in PDB 1ala may be indicative of positive selection acting to confer a beneficial structural conformation, although more evidence is required to substantiate this conclusion. \\

The presence of the identified jump in $11$ protein pairs with high posterior probabilities suggests that this may represent a common evolutionary motif. Conjecturally, the identification of evolutionary motifs together with improved modelling may in the future prove useful for several reasons: \textit{i}) improvement in homology modelling predictions due to the more accurate prediction of large conformational changes; \textit{ii}) improved estimates of evolutionary parameters, such as evolutionary times, which may be inflated when large conformation shifts are not modelled; and \textit{iii}) may help identify classes of functionally relevant positions that are potential drug targets, given that large changes in dihedral angles might be associated with consequential and predictable functional changes. 

\section{Conclusions}

We have shown that the WN process, an ergodic, time-reversible, and tractable Ornstein--Uhlenbeck toroidal analogue, is a cornerstone in the development of ETDBN, a tractable, generative, and interpretable probabilistic model of protein sequence and structure evolution on a local scale. The probabilistic nature of ETDBN allows rigorous statements about uncertainty to be made. The ability to infer various quantities of interest (such as evolutionary times or missing structures) and to interpret the parameters of the model demonstrates ETDBN's usefulness in gaining biological insights. Many existing computational models of biological structures lack statistical rigour, relying on heuristic techniques that do not provide any quantification of uncertainty. We envisage that the use of problem-adapted statistical methods, like the toroidal diffusions considered in this chapter, will grow in importance as practitioners increasingly demand a rigorous understanding of the underlying assumptions and inferences made in the construction and application of their models.

\section*{Acknowledgements}

This work is part of the Dynamical Systems Interdisciplinary Network, University of Copenhagen, and was funded by the University of Copenhagen 2016 Excellence Programme for Interdisciplinary Research (UCPH2016-DSIN), and by project MTM2016-76969-P from the Spanish Ministry of Economy, Industry and Competitiveness, and European Regional Development Fund (ERDF).


\end{document}